\documentclass[]{aastex631}

\newcommand{\fnum}{erg cm$^{-2}$ s$^{-1}$}

\shorttitle{AASTeX v6.31 Sample article}
\shortauthors{Kowalski}
\graphicspath{{./}{figures/}}

\begin{document}
\title{Bridging High-Density, Electron Beam Coronal Transport and Deep Chromospheric Heating in Stellar Flares}

\correspondingauthor{Adam F Kowalski}
\email{adam.f.kowalski@colorado.edu}

\author[0000-0001-7458-1176]{Adam F. Kowalski}
\affiliation{National Solar Observatory, University of Colorado Boulder, 3665 Discovery Drive, Boulder, CO 80303, USA}
\affiliation{Department of Astrophysical and Planetary Sciences, University of Colorado, Boulder, 2000 Colorado Ave, CO 80305, USA}
\affiliation{Laboratory for Atmospheric and Space Physics, University of Colorado Boulder, 3665 Discovery Drive, Boulder, CO 80303, USA.}



\begin{abstract}
The optical and near-ultraviolet (NUV) continuum radiation in M dwarf flares is thought to be the impulsive response of the lower stellar atmosphere to magnetic energy release and electron acceleration at coronal altitudes. This radiation is sometimes interpreted as evidence of a thermal photospheric spectrum with $T \approx 10^4$ K. However, calculations show that standard solar flare coronal electron beams lose their energy in a thick target of gas in the upper and middle chromosphere (log$_{10}$ column mass /[g cm$^{-2}$] $\lesssim -3$). At larger beam injection fluxes, electric fields and instabilities are expected to further inhibit propagation to low altitudes. We show that recent numerical solutions of the time-dependent equations governing the power-law electrons and background coronal plasma (Langmuir and ion-acoustic) waves from Kontar et al.\ produce order-of-magnitude larger heating rates than occur in the deep chromosphere through standard solar flare electron beam power-law distributions. We demonstrate that the redistribution of beam energy above $E \gtrsim 100$ keV in this theory results in a local heating maximum that is similar to a radiative-hydrodynamic model with a large, low-energy cutoff and a hard power-law index.  We use this semi-empirical forward modeling approach to produce opaque NUV and optical continua at gas temperatures $T \gtrsim 12,000$ K over the deep chromosphere with log$_{10}$ column mass /[g cm$^{-2}$] of $-1.2$ to $-2.3$.  These models explain the color temperatures and Balmer jump strengths in high-cadence M dwarf flare observations, and they clarify the relation among atmospheric, radiation, and optical color temperatures in stellar flares.

\end{abstract}



\section{Introduction} \label{sec:intro}

Empirical, multi-wavelength relationships in solar and stellar flares are consistent with similar physical processes of magnetic reconnection, particle acceleration, and atmospheric heating \citep{Hawley1995,Gudel1996, Neupert1968, Dennis1993}.  However, non-negligible optical depths from plasma at $T \approx 10,000$ K in the near-ultraviolet (NUV) and optical continuum are critical in reproducing M-dwarf spectral flare observations \citep{Livshits1981, HF92, Kowalski2015}, which are incompatible with most inferences from solar flare observations in the optical \citep{Potts2010} and NUV \citep{Heinzel2014, Kleint2016, Kowalski2017Mar29, Dominique2018}.  Lower rates of atmospheric excitation result in smaller electron densities and optical depths in solar flare chromospheres  \citep{Neidig1983, Kowalski2022}.  In solar flare chromospheric heating models, the only self-consistent predictions of optically thick flare continua originate from relatively small temperature increases, $\Delta T \approx 500-1000$ K in the radiatively backwarmed photosphere \citep[][ see also \citealt{Kleint2016}]{Neidig1993, Allred2005, Allred2006, Cheng2010, Kowalski2017Mar29}.  However, it is unclear whether solar spectra have yet sampled the brightest sources and whether there is more optically thick continuum radiation at certain very bright sources than models predict in solar flares.  There is a variety of spectra in the optical and $U$ band that have been reported in solar flares \citep{Neidig1983, Neidig1993, KCF15, Ondrej1}, which is evidence  that the comprehensive processes that heat the chromosphere and photosphere are not fully understood.

Large continuum optical depths in M dwarf flares have been achieved through static, semi-empirical modeling \citep{Cram1982, Christian2003, Fuhrmeister2010, Kowalski2011} and in radiative-hydrodynamic (RHD) modeling with extremely large electron beam flux densities\footnote{Hereafter, we abbreviate the beam flux density [erg cm$^{-2}$ s$^{-1}$] as ``beam flux''.} of $10^{13}$ \fnum\ distributed with a power-law  index, $\delta \approx 3-4$, above a low-energy cutoff of $E_c \approx 30-40$ keV  \citep{Kowalski2015}.  These particular electron beam parameters were inferred from collisional thick target modeling of hard X-ray and gamma ray spectra of the large 2002 July 23 X4.5 solar flare \citep{Holman2003, White2003, Allred2006, Ireland2013}.  The collisional thick target model  \citep{Brown1971} assumes that Coulomb collisions  dominate the energy loss  \citep[e.g.,][]{Emslie1978, Leach1981} of nonthermal electrons as they radiate hard X-rays in the thick target chromosphere (see \citealt{Brown2009} and \citealt{Kontar2011} for overviews).  It  is the most widely used framework \citep[e.g.,][]{Milligan2014, Kleint2016, Dennis2019} in forward modeling thermal spectra of solar flare RHD processes \citep[e.g.,][]{Allred2005, Rubio2016, Kowalski2017Mar29, Sadykov2019, Graham2020} and in hypothesis testing of X-ray spectra of stellar superflares \citep{Osten2007, Osten2016}. 
In recent higher spatial resolution solar observations of chromospheric/photospheric flare footpoint sources, \citet{Krucker2011} infer high electron beam fluxes of $10^{12} - 10^{13}$ \fnum\ above $18-20$ keV under the assumptions of the standard collisional thick target model. They interpret this beam flux range to mean that there is a severe, systematic flaw \citep[e.g.,][]{Smith1975, Brown1977} in the standard assumptions.  Magnetic field convergence \citep{Kontar2008} has been briefly mentioned as a potential solution \citep{Brown2009}, but \citet{Krucker2011} exclude this  explanation in their analysis.  In a limb flare, \cite{Martinez2012} directly imaged white-light and hard X-ray source heights and found lower altitudes than expected through state-of-the-art modeling of time-independent electron beam transport from a coronal source \citep[e.g.,][]{Battaglia2012}. 

Some radio observations of gyrosynchrotron emission also point to very large nonthermal electron densities of $n_e \approx 10^{10}$ cm$^{-3}$ in the corona \citep{White2003, Raulin2004, Kundu2009, White2011, Kawate2012}; these measurements are not dependent on the assumptions of the collisional thick target model for hard X-rays.   In the 2002 July 23 solar flare, \citet{White2003} discuss that the similarity between the radio light curves from the dense,  $n_e \approx 10^{10}$ cm$^{-3}$, coronal flare sources  and from the hard X-ray footpoint sources  implies origins from the same population of accelerated electrons.   Most recently, unprecedented observations from the Expanded Owens Valley Solar Array have shown that ambient particles evacuate into a nonthermal distribution over large coronal volumes \citep{Fleishman2022}.  These persistent sources are attributed to a trapping mechanism that is yet to be specified.

If the dense radio-emitting sources precipitate as large fluxes into the chromosphere  at the locations of the bright\footnote{Large energy fluxes into the chromosphere should also produce very bright emission lines and continuum intensity spectra at the locations of the brightest kernels.  To be consistent with current observational limits, very small filling factors  would be required.} solar flare kernels,  non-collisional transport physics must be considered.  The relative displacement of the accelerated electrons and protons generates a strong electric field that drives a cospatial return current (drifting Maxwellian) of the ambient electrons.  The neutralizing return current is also required to prevent enormous magnetic fields.  A large beam current density decelerates in this electric field and loses its energy to the background plasma through Joule heating \citep{Holman2012}.  Some analytic calculations suggest this may occur over just several hundred meters \citep{Oord1990}.  

It is generally thought that large beam densities cannot even form before steady state, or the beams should mostly confine themselves to the coronal acceleration region in the presence of electric field double layers that rapidly develop in response to large relative drift velocities between the ambient electrons and ions \citep[e.g.,][]{Li2012, Li2014}.  \citet{Lee2008} simulate  a very hot Maxwellian beam and calculate the energy loss due to plasma instabilities.  They suggest that the highest-energy electrons in the beam propagate to the footpoints with less energy loss than the lower energy particles, which thermalize.  For a power-law distribution, however, there are relatively few electrons in the high-energy tail to begin with.  Other calculations predict very large amounts of energy loss, $\approx 0.6 m_i c_{\rm{Alfven}}^2$, during beam passage through a series of double-layer electric fields \citep{Li2014}.     Thus, hard X-ray footpoint sources would be dominated by thermal emission if the bulk of beam energy thermalizes during coronal transport, which is contrary to their well-understood properties that include nonthermal electron bremsstrahlung radiation at $E \approx 25-300$ keV, gamma-ray nuclear excitation lines, and electron-positron annihilation radiation \citep{Vilmer2011, Dennis2022}.  Moreover, beam energy loss in the corona and subsequent thermal conduction into the chromosphere produce faint continuum radiation \citep{Kowalski2017Mar29} and are not able  to explain large optical depths and hydrogen line broadening in M dwarf flare observations \citep{Namekata2020}.  On the other hand, updated hydrogen pressure broadening in the RHD models of \citet{Kowalski2015} and \citet{Kowalski2016} suggest that the ``scaled-up'' beam fluxes that are characterized by power-law parameters, as inferred from solar flare hard X-rays, produce chromospheric condensations that are far too dense to be consistent with optical stellar spectra \citep{Kowalski2017Broadening, Kowalski2022Frontiers}.

An alternative heating mechanism to extreme-flux, electron-beam distributions that are inferred from standard collisional thick target modeling of solar flares  is warranted to explain the deep chromospheric heating in M dwarf flares while also accounting for transport effects due to large coronal beam densities.  This topic is timely in the context of the many thousands of white-light stellar flares that have recently been reported in data from Kepler, K2, and TESS \citep[e.g.,][]{Hawley2014, Maehara2021}. The potential impact of ultraviolet flares is also developing into an important issue facing assessments of exoplanet habitability \citep[e.g.,][]{Loyd2018, Howard2018, Tilley2019, Howard2020, Abrevaya2020}.   One such alternative modeling approach is explored in  \citet{Kowalski2017Broadening}, where they extend the standard electron beam modeling paradigm to larger, low-energy cutoff values ($E_c = 85-500$ keV) than typically inferred ($E_c \approx 15-25$ keV) in solar flares, aside from a few instances reported in late impulsive peaks \citep{Holman2003, Warmuth2009}.   In \citet{Kowalski2022Frontiers}, we develop this modeling approach further for comparisons to the entire observed hydrogen Balmer line series in stellar flare spectra from \citet{HP91}, and we achieve statistical agreement with the rise and peak phase.  However, physical justification for large, low-energy cutoffs of accelerated electron distributions in M dwarf flares has not been formulated.
Here, we report on extensions to the modifications of the collisional thick target model that are developed in \citet{Kontar2012}; hereafter K12.  We show that the time-dependent, coronal transport calculations of K12 produce  beam heating distributions in the pre-flare, low chromosphere that are remarkably similar to electron beams with large, low-energy cutoffs ($E_c \approx 85$ keV) and hard ($\delta \approx 3$) power-law indices.


The paper is organized as follows. 
In Section \ref{sec:setup}, we describe the RHD flare model setup.  We analyze several RHD flare models with large, low-energy cutoffs that generate optical depths in the NUV and optical continuum in the low chromosphere (Section \ref{sec:optdepth}).  We compare the highest-flux model to high-cadence optical flare colors that were reported recently in the literature (Section \ref{sec:ultracam}).    In Section \ref{sec:langmuir}, we present new calculations of the initial chromospheric heating profiles using the time-averaged electron beam spectra in K12, and we compare them to the models that provide new interpretations of the flare data.   In Section \ref{sec:discussion}, we discuss the implications for future modeling efforts of solar and stellar flares.   In Section \ref{sec:conclusions}, the main conclusions of this work are presented.

\section{RHD Flare Modeling with RADYN} \label{sec:rhd}

Before presenting heating rates from the modified beam distribution of K12, we establish that deep atmospheric heating from a large, low-energy cutoff electron beam model and an extremely high energy flux produce large continuum optical depths at $\lambda = 3615$ \AA, $4170$ \AA, and $6010$ \AA.  We analyze the formation of the continuum radiation at these wavelengths,  and we explain  how hot, optical color temperatures ($T_{\rm{col}} \approx 10^4$ K) in spectral observations originate in these models.  To establish the importance of the K12 theory for this phenomenological characteristic of stellar flares, we compare to constraints from high-cadence flare colors during the rise and peak phases throughout a giant dMe flare event that was reported in \citet{Kowalski2016}.

\subsection{Electron Beam Heating with Large Low-energy Cutoffs:  Model Setup}  \label{sec:setup}
We have calculated a comprehensive grid of RHD models with the \texttt{RADYN} code \citep{Carlsson1992B, Carlsson1995, Carlsson1997, Carlsson2002, Allred2015}.  The details about the setup will be described in a separate paper (Kowalski et al.\ 2023, in preparation),  but a brief summary is presented here.  The effective temperature of the starting atmosphere is $T_{\rm{eff}} \approx 3600$ K \citep[see the Appendix of][for details regarding the starting atmosphere]{Kowalski2017Broadening}.  The equations of mass, momentum, internal energy,  and charge are solved on an adaptive grid \citep{Dorfi1987} with the equations of radiative transfer and level populations for hydrogen, helium, and Ca II. 
To simulate flare heating, we model the energy deposition from a power-law distribution of electrons, which is calculated in a 1D magnetic loop of half-length $10^9$ cm, a constant surface gravity of log\ $g=4.75$, and a uniform cross-sectional area.  The electron beam is injected at the loop apex, which has an ambient electron density ($n_e$) of $3 \times 10^{10}$ cm$^{-3}$ and a gas temperature ($T_{\rm{gas}}$) of 5 MK.  The atmospheric response is calculated with a ramping beam flux to a maximum value at $t=1$~s, followed by a decrease until $t=10$~s according to the pulsed injection profile prescription in \citet{Aschwanden2004}. We assume that the injected pitch angle distribution is Gaussian-distributed in $\mu = \cos \theta$ (where $\mu = 1$ is directed along the magnetic loop axis) in the forward hemisphere with a spread of $\sigma_{\mu} = 0.07$.
 The heating rates as a function of atmospheric depth at each time step in the RHD simulations are calculated from the steady-state solution of the Fokker-Planck equation for Coulomb energy loss to neutrals and charged particle constituents; this solver was recently developed into the \texttt{FP} code \citep{Allred2020}.  In this first-generation M dwarf flare model grid, return current and magnetic mirroring forces are not included as external force terms.  Thus, these RHD simulations capture the physics of ``free-streaming'' electron beam propagation. In general\footnote{The RHD response also includes time dependence and detailed atmospheric variations of ionization.} terms, these models  are consistent with the assumptions of the collisional thick target model of hard X-rays.
 The peak injected beam energy fluxes span  four orders of magnitude: $10^{10}$ (F10), $10^{11}$ (F11), $10^{12}$ (F12), and $10^{13}$ (F13) erg cm$^{-2}$ s$^{-1}$.  The low-energy cutoffs in the grid range from $E_c = 17$ keV to 500 keV.   The selected pulsed injection models for this study have $E_c = 85$ keV and injected electron beam number fluxes with a  power-law index of $\delta = 3$, which is consistent with available stellar flare constraints \citep{Osten2007, MacGregor2018, MacGregor2020, MacGregor2021}.  We refer to these models as mF10-85-3, mF11-85-3, mF12-85-3, and mF13-85-3.  Following \citet{Kowalski2017Mar29} and the appendices of \citet{Kowalski2017Broadening}, the contribution functions ($C_I$) to the emergent intensity, the optical depths ($\tau$), and the radiation (brightness) temperatures ($T_{\rm{rad}}$) at several continuum wavelengths are analyzed at a viewing angle corresponding to $\mu^{\prime} = \cos \theta^{\prime} = 0.95$ at a time of $t=1$~s.    This time corresponds to the maximum beam energy injection into the atmosphere.
A corresponding grid of models is calculated using a constant (``c'') beam flux injection; the initial heating rate at $t=0$~s for the constant injection model, cF13-85-3, is discussed in Section \ref{sec:langmuir}.  The optical depth calculations in Section \ref{sec:optdepth} are nearly identical for the constant flux injection models at $t = 1$~s; therefore, we choose to analyze in detail the ramping beam injection calculations that modeled low-time resolution spectral data of a dMe flare in \cite{Kowalski2022Frontiers}.

\subsection{Model Continuum Spectrum Analysis} \label{sec:optdepth}

\begin{figure}
\gridline{\fig{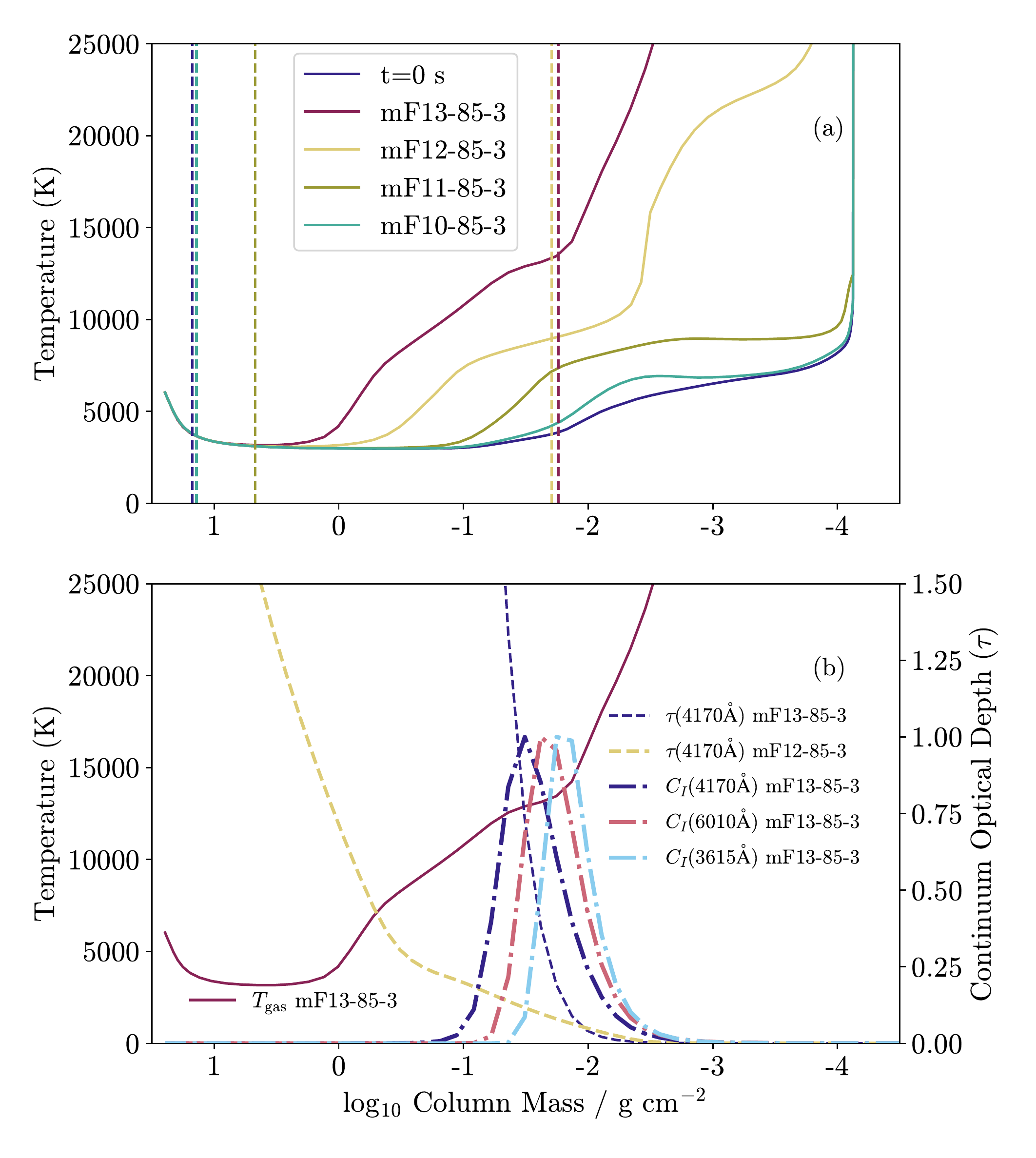}{0.6\textwidth}{}}
\caption{ \textbf{(a)} The atmospheric gas temperature response to electron beam-heating models with $E_c = 85$ keV at $t=1$~s after injection into a model M dwarf atmosphere.  The vertical dashed lines indicate the column masses at which $\tau =0.95$ occur in the continuum wavelength at $\lambda = 3615$ \AA.  The preflare gas temperature ($t=0$~s) is shown for comparison.  \textbf{(b)}    The variation of the optical depth at $\lambda = 4170$ \AA\ exhibits marked differences between the mF12-85-3 and mF13-85-3 models at $t=1$~s in their respective evolution. The gas temperature of the mF13-85-3 model is reproduced from panel (a).   The contribution function ($C_I$) to the emergent intensity at three continuum wavelengths in the mF13-85-3 model at $t=1$~s is displayed in units of erg cm$^{-2}$ s$^{-1}$ \AA$^{-1}$ s.r.$^{-1}$ (unit log$_{10}$ col mass)$^{-1}$ on a linear scale with arbitrary peak normalization.
\label{fig:optical_depths}}
\end{figure}

The range of maximum beam fluxes from the mF10-85-3 model to the mF13-85-3 model establishes a threshold at which non-negligible continuum optical depths develop over the low atmosphere as a function of wavelength.
Figure \ref{fig:optical_depths}(a) shows the gas temperature response of the M dwarf atmosphere at $t=1$~s in each model.  The vertical dashed lines indicate the locations of $\tau=0.95$ at the continuum wavelength of $\lambda = 3615$ \AA, which is a representative wavelength on the blue side of the Balmer jump.  As the maximum beam flux of the model increases to the F13 level, the effective photosphere at this wavelength shifts up to the column mass (log$_{10}\ m$ / [g cm$^{-2}$]) range between $\approx -1.7$ and $-2.0$, which corresponds to the lower chromosphere in the preflare state.  In the mF13-85-3, the $\tau \approx 1$ layer occurs at $T_{\rm{gas}} \approx 13,500$ K.    The  NUV flare photosphere occurs at a cooler temperature, $T_{\rm{gas}} \approx 9000$ K, in the mF12-85-3 simulation, but the optical depth variation at $\lambda = 3615$ \AA\ as a function of column mass (not shown) is rather similar to the mF13-85-3.

In Figure \ref{fig:optical_depths}(b), the mF13-85-3 and mF12-85-3 models clearly differ in the optical depths at longer continuum wavelengths.  The variation of $\tau$ at $\lambda = 4170$ \AA\  (shown for the mF13-85-3 and mF12-85-3 models)  confirms that only the mF13-85-3 model produces significant optical depth in the significantly heated regions of the deep chromosphere around log$_{10}\ m \approx -1.5$, which becomes the optical flare photosphere at $t=1$~s.    The contribution functions to the emergent intensity at $\lambda=4170$ \AA, $\lambda = 6010$ \AA, and $\lambda = 3615$ \AA\ are shown from the mF13-85-3 model in Figure \ref{fig:optical_depths}(b).  The continuum at $\lambda = 4170$ \AA\ is formed over a deeper range of column mass at $t=1$~s than the NUV continuum intensity at $\lambda = 3615$ \AA\ and the red-optical continuum intensity at $\lambda = 6010$ \AA.   Among all NUV, optical, and near-infrared continuum wavelengths, the blue-optical wavelengths at $\lambda \approx 4000-4200$ \AA\ are the most optically thin at $T_{\rm{gas}} \approx 10^4$ K.  Note that of the semi-empirical, static models in \citet{Cram1982}, their model \# 5 is most similar to the mF13-85-3 atmospheric state at $t=1$~s.     In the appendices of \citet{Kowalski2017Broadening}, similar contribution functions at $\lambda = 4170$ \AA\  for two models with lower beam fluxes and larger, low-energy cutoffs are described.  In the next section, we connect the multi-wavelength continuum formation from the mF13-85-3 model to the broadband continuum shape in the impulsive phase of a large M dwarf flare.

\subsubsection{Comparison to High-cadence Flare Color Observations} \label{sec:ultracam}
\begin{figure}
\begin{center}
\includegraphics[scale=0.5]{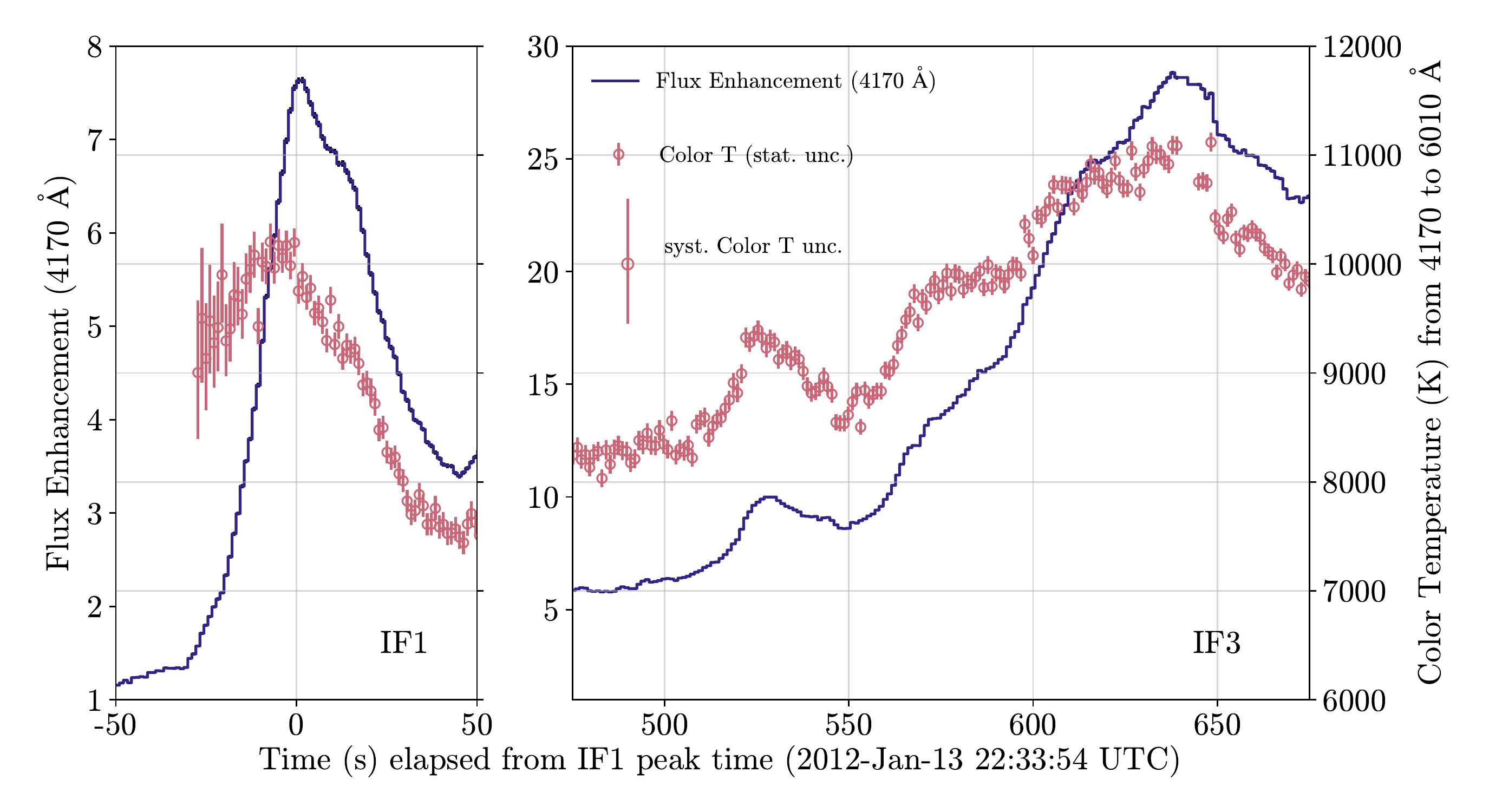}
\caption{   High-cadence ($\Delta t = 1.1$~s) ULTRACAM photometry in the NBF4170 ($\lambda_{\rm{cen}} = 4170$ \AA) narrowband continuum filter during two sequential flare events (IF1 and IF3) on the dM4.5e star YZ CMi. The photometry is normalized to 1.0 during quiescence, and the statistical photometry errors are plotted but are small on this scale.  The flux-calibrated, quiescent-subtracted NBF4170 and the RC\#1 ($\lambda_{\rm{cen}} = 6010$ \AA) data are converted to filter ratios.   We solve for the optical color temperatures ($T_{\rm{col}}$), which are referred to as FcolorR in the study of these events in \citealt{Kowalski2016}.  Representative systematic uncertainties on the color temperatures are indicated as $^{+600}_{-550}$ K and are obtained from the uncertainties in the absolute flux calibration.  Note that detailed comparisons of spectra to ULTRACAM filter ratios in smaller-amplitude flares that occurred later in these observations have revealed that values of TFcolorR $\approx 10,000$ K are systematically smaller by $\approx 2000$ K compared to the color temperatures that are fit to the blue-optical spectral range at $\lambda = 4000-4800$ \AA\  \citep[see][for details about the data, flare colors, and error analysis]{Kowalski2016}.}
\label{fig:yzcmi_flare}
\end{center}
\end{figure}

The emergent radiative flux and intensity predictions from the mF13-85-3 model are consistent with spectral observations of the impulsive phase of some M dwarf flares.  Specifically, the measured spectral quantities that motivate large heating rates are small Balmer jump ratios and hot, NUV, blue-optical, and red-optical color temperatures of $T_{\rm{col}} \approx 10,000 - 11,000$ K  \citep{Mochnacki1980, HP91, Fuhrmeister2008, Kowalski2013, Kowalski2016, Kowalski2018, Kowalski2022Frontiers}.   Two sequential, high-energy flare events that occurred on the M-dwarf YZ CMi are shown in Figure \ref{fig:yzcmi_flare}.  These events were observed in three narrowband continuum filters with central wavelengths at $\lambda_{\rm{cen}} = 3500$ \AA, $4170$ \AA, and $6010$ \AA\ using the ULTRACAM instrument \citep{Dhillon2007}.  These flares were studied in detail in \citet{Kowalski2016}, and are referred to as the ``IF1'' and ``IF3'' events in that work.    Following  the method of \citet{Hawley1995}, we solve for blackbody color temperatures through Newton-Raphson linearization and iteration constrained to the $\lambda_{\rm{cen}} = 4170$ \AA\ and $\lambda_{\rm{cen}} = 6010$ \AA\ continuum-filter ratios\footnote{Data and flare-only filter ratios are publicly available from Zenodo; \url{https://doi.org/10.5281/zenodo.45878
}} that are reported in \citet{Kowalski2016}.  

Broadband, white-light color temperatures of $T_{\rm{col}} = 9,000 - 11,000$ are attained over the rise, peak, and initial fast-decay phases of these events.    The flares in Figure \ref{fig:yzcmi_flare} exhibit the highest-time resolution constraints on the broadband color temperature evolution in a dMe flare event.  Similar hot, blue-optical color temperatures were calculated from spectra during the rise, peak, and fast decay phase during another large-amplitude event on YZ CMi in \cite{Kowalski2013}, and a detailed comparison of hot color temperatures from simultaneous spectra and ULTRACAM photometry were analyzed in smaller-amplitude events after IF3 in \cite{Kowalski2016}.  The evolution of the color temperature to cooler values after each major peak in Figure \ref{fig:yzcmi_flare} has been reported in spectra of the gradual decay phases of other large dMe flares \citep[see Fig. 31 of ][]{Kowalski2013}, which indicate that a single-continuum model is unable to explain the full optical continuum distribution.  Here, we focus on the hotter phases of the IF1 and IF3 events, which have long-challenged models with lower flux electron beams and X-ray backwarming  \citep[e.g.,][]{HF92, Allred2006}.  The phenomenological ``9000 K blackbody hypothesis'' is widely adopted to facilitate calculating flare energies and ultraviolet fluxes using extrapolations from single-bandpass Kepler, K2, and TESS optical/infrared photometry \citep[e.g.,][]{Shibayama2013, Gunther2020, Chen2021}. A self-consistent physical explanation, as follows, of the relatively short but luminous phases around each major peak in Figure \ref{fig:yzcmi_flare} is of broad significant astrophysical interest.

Thus, we further analyze the emergent continuum intensity, radiation temperature, and  contribution function dependencies on wavelength from the mF13-85-3 model at $t=1$~s in order to explain the hottest values of $T_{\rm{col}} \approx 10,000 - 11,000$ K in the ULTRACAM data of the  YZ CMi flare events in Figure \ref{fig:yzcmi_flare}. 
 The continuum  intensity at $\lambda = 4170$ \AA\ is more optically thin and thus forms over slightly deeper, cooler temperatures (Figure \ref{fig:optical_depths}b) than at $\lambda = 6010$ \AA.  Consequently, the radiation temperatures of the model spectra at $\lambda = 4170$ \AA\ and $\lambda = 6010$ \AA\ are $T_{\rm{rad}} = 13,200$ K and $14,100$ K, respectively.   From the respective emergent continuum intensity ratios, we solve for a color temperature of $T_{\rm{col}} = 10,900$ K, which is consistent\footnote{The intensity spectrum at $t=1$~s provides an upper limit to the model prediction.  It is outside the scope of this work to consider statistical averages over ULTRACAM exposure times and heterogeneous stellar flare flux sources; we estimate that these considerations would lower the color temperatures predicted by this model by only several hundred degrees; see \citet{Kowalski2022Frontiers}.} with the color temperatures from the ULTRACAM filter ratios (Figure \ref{fig:yzcmi_flare}).  However, the model $T_{\rm{col}}$ is below the atmospheric gas temperature range,  $T_{\rm{gas}} \approx 12,000 - 20,000 $ K, over which 90\% of the emergent optical and NUV continuum intensity originates.   This hot and heterogeneously stratified atmospheric temperature structure is fully consistent with a color temperature in the emergent spectrum that is apparently cooler and isothermal, $T_{\rm{col}} \approx 9000 - 11,000$ K.   We find the following approximate relationship among these temperature measures:  $T_{\rm{rad}} \approx T_{\rm{gas}} (\tau \approx 0.3) > T_{\rm{col}}$.    Thus, color temperatures calculated from spectra are not direct measures of the atmospheric gas temperature in these model atmospheres that exhibit large optical depths in the deep atmosphere.
The continuum formation is near local thermodynamic equilibrium (LTE), but the Eddington-Barbier relation is not an accurate simplification of the continuum formation\footnote{We confirm this by analyzing the dependence of the non-LTE source function with $\tau$ over regions of the atmosphere with significant values of the contribution function.} because $T_{\rm{rad}} \ne T_{\rm{gas}}(\tau=0.95)$.   Similar analyses explain the small model Balmer jump ratio, which is consistent with the measured ratios of the $\lambda \approx 3500$ \AA\ to $\lambda \approx 4170$ \AA\ fluxes \citep[see][]{Kowalski2016} over the times shown in Figure \ref{fig:yzcmi_flare}. 
 
\subsection{Electron-beam Generated Langmuir and Ion-acoustic Turbulence} \label{sec:langmuir}
Electron beams with high-energy fluxes ($\gtrsim 10^{12}$ \fnum) and large low-energy cutoffs ($E_c \gtrsim 80$ keV) are therefore well-justified, semi-empirical models for the heating rates that generate continuum optical depths in the low atmosphere during M-dwarf flares \citep[see also][]{Kowalski2022Frontiers} at very high-time resolution.  There are many implications for models of the Balmer jump, NUV, and Balmer line broadening that are outside the scope of this paper but will be presented in future work.   These applications include models of less impulsive flare types that exhibit larger Balmer jumps and smaller optical color temperatures \citep{Kowalski2019HST}, flares that exhibit Balmer jumps in absorption and hotter optical color temperatures \citep{Kowalski2017Broadening} than the events in Figure \ref{fig:yzcmi_flare}, and the evolution of the flare colors through the gradual decay phase (e.g., in Figure \ref{fig:yzcmi_flare}).   We now turn to the main result of this work in which we propose the physical origin for  stellar flare electron beams with effective low-energy cutoffs of $E_c \gtrsim 85$ keV and hard power-law indices.  This result draws on and connects to theoretical foundations that have been recently developed to modify the collisional thick target model of solar flare hard X-ray emission.

The propagation of electron beams on very short timescales, $\Delta t \ll 1$~s, after the initial acceleration process(es) has been investigated with collisionless (Vlasov-Maxwell), particle-in-cell (PIC) simulations \citep[see][for modern reviews and discussions of recent progress]{Arber2015, Nishikawa2021}.  Alternatively, numerical solutions over longer timescales are possible for the set of coupled equations consisting of a collisional, time-dependent equation that governs the electron (beam$+$plasma) phase-space evolution and an equation that governs the evolution of background plasma wave energy (K12; see also \citealt{Hamilton1987, Kontar2001, Kontar2002, Hannah2009, Hannah2013, Ratcliffe2014, Thorne2017}).  The component of the distribution function for the background plasma represents the evolution of longitudinal plasma waves, which consist of ambient electron disturbances (Langmuir waves) and ion sound waves (ion-acoustic waves).  K12 solve these equations over $\Delta t = 1$~s of electron beam propagation through the solar corona with random (Gaussian) density perturbations simultaneously with non-linear, three-wave interactions \citep[e.g.,][and Ch.\ 23.3.6 of \citealt{Thorne2017}]{Tsytovich1995} and a term for collisional energy loss from the beam particles.   To briefly summarize the results that are relevant to this work, the kinetic energy losses of low-energy beam electrons generate Langmuir waves, which evolve in angular wavenumber ($k$) space through wave-wave processes (and through refraction and diffusion).  The wave turbulence transfers energy back to the beam, resulting in effective energy gains of $dE/dt \gtrsim 200$ keV s$^{-1}$ per beam electron at $E \approx  100-400$ keV (see Figs.\ 4 -- 5 of K12).

\begin{figure}
\gridline{\fig{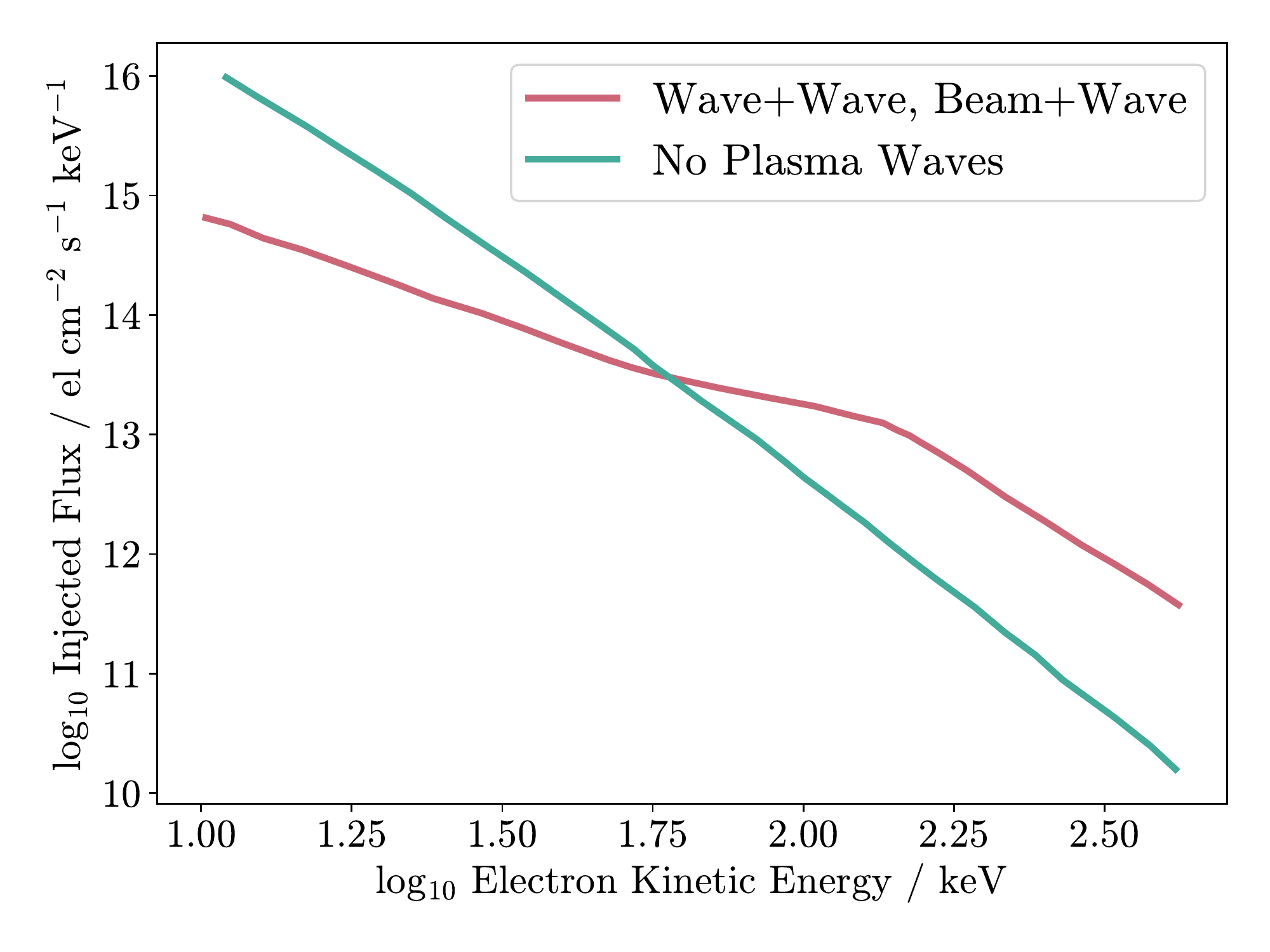}{0.38\textwidth}{(a)}
         \fig{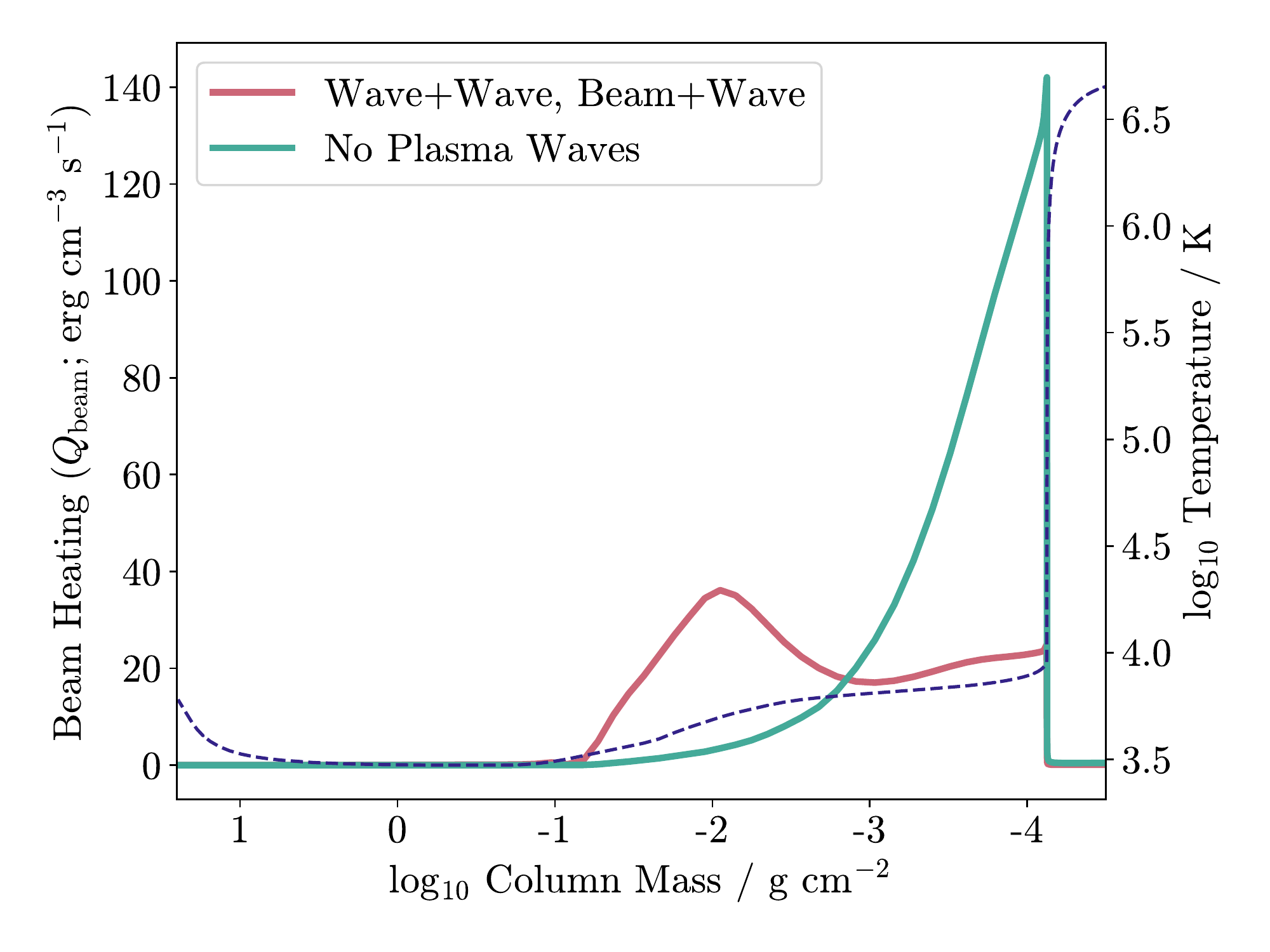}{0.4\textwidth}{(b)}
         }
\gridline{\fig{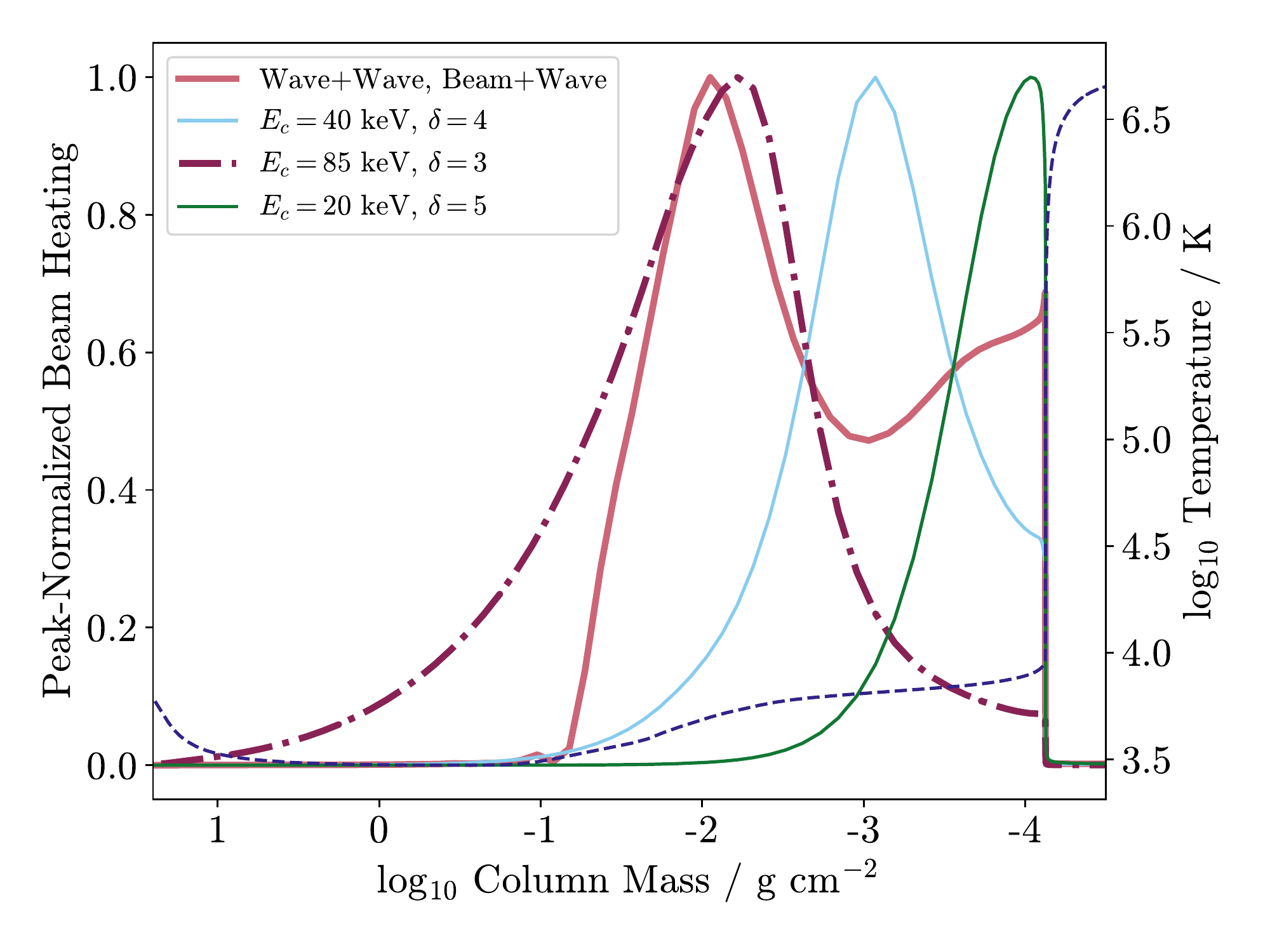}{0.6\textwidth}{(c)}
\fig{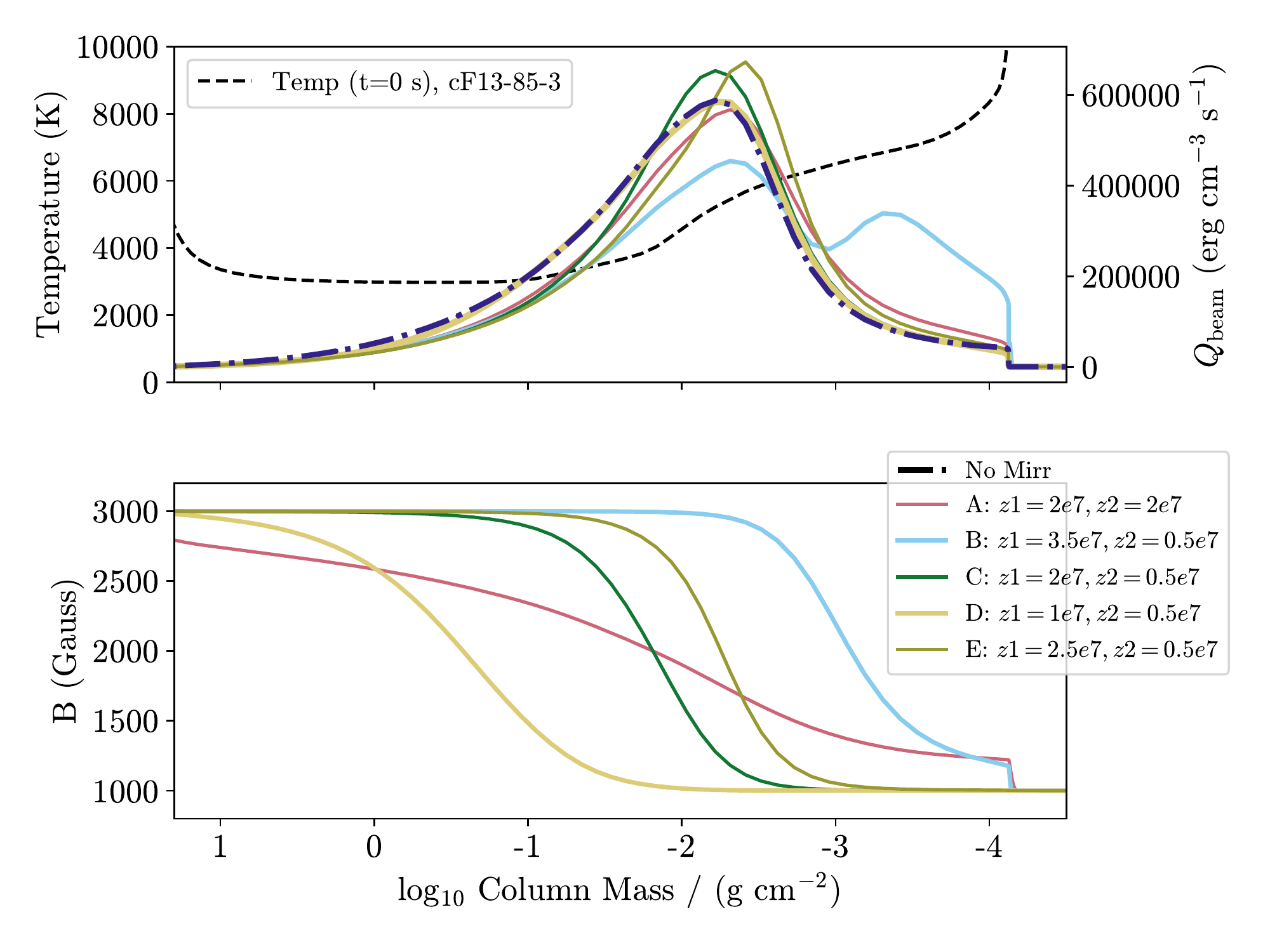}{0.5\textwidth}{(d)}
}
\caption{ \textbf{(a)}  Nonthermal electron number flux distribution and time-averaged electron number flux distribution with plasma wave interactions, reproduced from Fig.\ 4 of K12.  \textbf{(b)} The initial heating rate from Coulomb collisions in  our \texttt{RADYN} M dwarf model chromosphere.    \textbf{(c)}  The initial Coulomb heating rate in the M dwarf chromosphere for several power-law indices and low-energy cutoffs compared to the K12 heating rate in panel (b).  (d) Beam-heating calculations (top panel) for various magnetic field convergence distributions in the chromosphere (bottom panel).  The initial gas temperature stratification of the lower M dwarf atmosphere is shown as dashed curves in panels (b)-(d).
\label{fig:beamdep}}
\end{figure}

We demonstrate that the enhancements of the numbers of beam electrons with $E \gtrsim 100$ keV at the expense of the numbers at $E \lesssim 60$ keV as a result of the energy transfer mechanisms in K12 effectively produce an energy deposition peak in the low chromosphere that is similar to the large, low-energy cutoff beams in Section \ref{sec:rhd}.   The time-averaged electron beam flux spectra (reproduced in Figure \ref{fig:beamdep}(a) here) from Fig.\ 4 of K12 are injected into a model M dwarf atmosphere from Section \ref{sec:rhd} to quantify the beam heating in a realistic model stellar chromosphere.  Hereafter, we refer to the beam flux spectrum from K12 that includes three-wave non-linear processes as the ``wave$+$wave, beam$+$wave'' simulation\footnote{For a lack of better shorthand notation, we use $+$ but do not intend for this to represent a linear superposition of wave amplitudes.}.   We solve the steady-state energy deposition with the Fokker-Planck module\footnote{We find similar general properties as these Fokker-Planck solutions using a simple numerical integrations of the energy losses given by the formulae and Coulomb logarithms in \citet{Emslie1978} and \citet{HF94}.} from  \citet{McTiernan1990} that was incorporated into \texttt{RADYN} for the flare simulations in \citet{Allred2015}.  This version of the \texttt{RADYN} code uniquely includes the capability for an arbitrary particle distribution function to be injected at the loop apex. We assume the same initial Gaussian distribution of pitch angle as for the large $E_c$ calculations  to facilitate comparison (Section \ref{sec:setup}).  The heating profile ($Q_{\rm{beam}}$) for the ``wave$+$wave,beam$+$wave'' model from K12 is shown for the starting \texttt{RADYN} M dwarf atmosphere in hydrostatic equilibrium in Figure \ref{fig:beamdep}(b).  Compared to the simulation without ``wave$+$wave, beam$+$wave'' interactions, the location of the maximum beam heating rate is shifted from the top of the chromosphere to the low chromospheric layers with a large column mass of log$_{10}\ m \approx -2$.  The ``wave$+$wave, beam$+$wave'' heating rate is an order of magnitude  larger at log$_{10}\ m \approx -2$ than the heating rate without plasma wave-beam interactions. 
This column mass corresponds to the collisional stopping depths of electrons with initial kinetic energies of $E \approx 100$ keV \citep{Emslie1978, HF94}. 

In Section \ref{sec:rhd}, we showed the importance of large amounts of heating over the column mass range of log$_{10}\ m$ from  $\approx -1.2$ to $-2.3$ in producing optically thick  continuum radiation at NUV and optical wavelengths.  We use the cF13-85-3 model calculation at $t=0$~s to show the initial beam heating distribution from a large, low-energy cutoff model in Figure \ref{fig:beamdep}(c) for direct comparison to the K12 heating prediction in the undisturbed M dwarf chromosphere.  The heating rates are normalized to their respective maximum values of $Q_{\rm{beam}}$.  The maximum remarkably corresponds to the same column mass range as the K12 ``wave$+$wave, beam$+$wave'' calculation.  We evaluate the fraction of beam flux lost to the column mass range between log$_{10}\ m = -2.3$ and $-1.2$ to quantify the similarity.  The cF13-85-3 and the K12 models result in fractions of 0.51 to 0.56, whereas standard beam distributions in solar and stellar modeling exhibit much smaller cumulative fractions.  A distribution with $E_c =  40$ keV represents a beam with about the largest cutoff value that is consistent with standard, collisional (cold) thick target modeling of solar flare hard X-rays \citep[e.g.][]{Holman2003,Ireland2013}; this beam deposits only 0.13 of its energy over the deep chromosphere.  More typically, values of $E_c = 20$ keV or less are inferred;  a beam with $\delta =5$ deposits less than 1\% of its integrated energy over this column mass range that is important for optically thick continuum formation at $T_{\rm{gas}} \gtrsim 10^4$ K.

\section{Discussion} \label{sec:discussion}
In the absence of mass advection, heating the chromosphere to $T_{\rm{gas}} \gtrsim 10^4$ K around log$_{10}\ m \approx -3$ results in blue continuum radiation that is much more optically thin than heating the chromosphere  around log$_{10}\ m \approx -1.5$ (\emph{cf.} Figure \ref{fig:optical_depths}(b)) to a comparable temperature.
The column mass range log$_{10}\ m$ of $-2.3$ to $-1.2$ corresponds to the flare layers that have been previously described as  ``stationary chromospheric flare layers'', which lie just below the chromospheric condensations in RHD simulations \citep[e.g.,][]{Kowalski2015}.  In the large, low-energy cutoff models, upflows develop ($5-20$ km s$^{-1}$) due to the thermal pressure gradients over this column mass range, where gas densities deviate from hydrostatic equilibrium at $t \gtrsim 5 $~s.  The modified electron beam distributions from K12 produce the relative enhancement in the high energy electrons $E \gtrsim 100$ keV that are needed to deliver a significant amount of released magnetic energy to these layers if the total energy flux in the beam is large enough.  The models with large low-energy cutoffs ($E_c \gtrsim 85$ keV, $\delta \approx 3$) are adequate approximations to the expected RHD response of the K12 beam in deep regions of the atmosphere.

 In future work, it will be important to also consider the upper atmospheric evolution in response to the injection of K12 electron beams.  Based on results from previous RHD models, we can make several qualitative predictions. 
We expect the low-energy electrons to readily increase the temperature of the corona and transition region through direct  collisional heating and through the steady-state return current heating \citep{Allred2020}, which has not yet been included in M dwarf flare models.
We expect these energy loss mechanisms to drive chromospheric condensations \citep[e.g.,][]{Graham2020}, which may build up continuum optical depths as they cool to $T_{\rm{gas}} \approx 10^4$ K.  Even without return current force terms, the larger Coulomb heating rates in the upper chromosphere of the K12 prediction in Figure \ref{fig:beamdep}(c) would likely generate chromospheric condensations that are not present in the beams with the larger low-energy cutoffs ($E_c \ge 85$ keV, $\delta=3$).
Furthermore, rapid thermal ionization of the chromosphere causes the beam heating to shift higher up within the chromosphere, and evaporation of chromospheric material eventually stops the low-energy particles in the low flare corona.  Magnetic field convergence in the low atmosphere is an additional external force that may shift the beam heating farther up in the chromosphere.

To evaluate the degree to which magnetic field convergence affects large beam fluxes at deep column masses around log$_{10}\ m \approx -2$, we show several $t=0$~s calculations in Figure \ref{fig:beamdep}(d) from the steady-state Fokker-Planck solution with magnetic mirroring of the particles, as described in \cite{Allred2020}.  We simulate the cF13-85-3 beam, but we widen the initial pitch angle distribution to have a spread of $\sigma_{\mu} = 0.18$ in order to accentuate the effects of pitch angle changes due to magnetic mirroring.  We follow \cite{Battaglia2012} and model the magnetic field using a hyperbolic tangent:  

\begin{equation}
B(z) = 2000 + 1000 \tanh \big(-\frac{z - z1}{z2} \big)
\end{equation}

\noindent with $z1$ and $z2$ indicated in the figure and $z$ the distance along the loop from $\tau_{500}=1$.  Even a carefully placed magnetic wall (convergence model E) does not largely affect the large heating rate maximum in the deep chromosphere. A convergence higher in the chromosphere (model B) still results in a very large heating maximum that is not predicted by standard electron beam distributions inferred through collisional thick target modeling.  The upper chromospheric heating more closely approaches the K12 heating rate (Figure \ref{fig:beamdep}(b)) in this region, while retaining a large flux into lower altitudes.  In convergence models A and B, we expect chromospheric condensations to develop with large heating rates below in the stationary flare layers.

The issue of timescales is possibly a more serious concern in the application of the K12 theory to the heating in a lower flaring atmosphere.  The times ($\Delta t= 1$~s) over which the K12 flux spectrum (reproduced in Figure \ref{fig:beamdep}(a)) is averaged are not consistent with the times of flight, which are on the order of $50-100$ ms, for mildly relativistic electrons to reach the chromosphere.  From the calculation in K12, it is not completely evident whether there is enough time for the ``wave$+$wave,beam+wave'' processes to operate in a wide variety of stellar atmosphere conditions.  However, the evolution begins with the onset of the fundamental (bump-on-tail) beam-plasma instability due to collisional loss of the lowest energy electrons, and the timescale for this  is faster\footnote{See, e.g., Ch.\ 2.7 of \cite{Tsytovich1995}, which derives the growth rate as $\gamma \approx \omega_{\rm{pe}} \frac{n_{\rm{beam}}}{n_{\rm{e}}} \frac{v_{\rm{beam}}^2}{(\delta v_{\rm{beam}})^2}$, where $v_{\rm{beam}}$ is the average beam velocity, $\delta v_{\rm{beam}}$ is its spread, and $\omega_{\rm{pe}}$ is the electron plasma (Langmuir) frequency, $\sqrt{\frac{4\pi n_e e^2}{m_e}}$; $n_e = n_{\rm{background}}$.  In \cite{Ratcliffe2012}, this is called the quasi-linear time and is investigated in detailed numerical simulations.} for larger values of $n_{\rm{beam}}/n_{\rm{background}}$, as expected for high-flux beams in M dwarf coronae; the ambient conditions of M dwarf coronae are briefly discussed below.  We also note that the energy transfer processes begin rapidly at $t \ll 1$~s in the calculation of K12 (see the upper right panel of their Fig.\ 4), but additional calculations with a range of $n_{\rm{beam}} / n_{\rm{background}}$ and temporal averages are probably warranted for input into RHD models of the chromospheric response.

\begin{figure}
\gridline{\fig{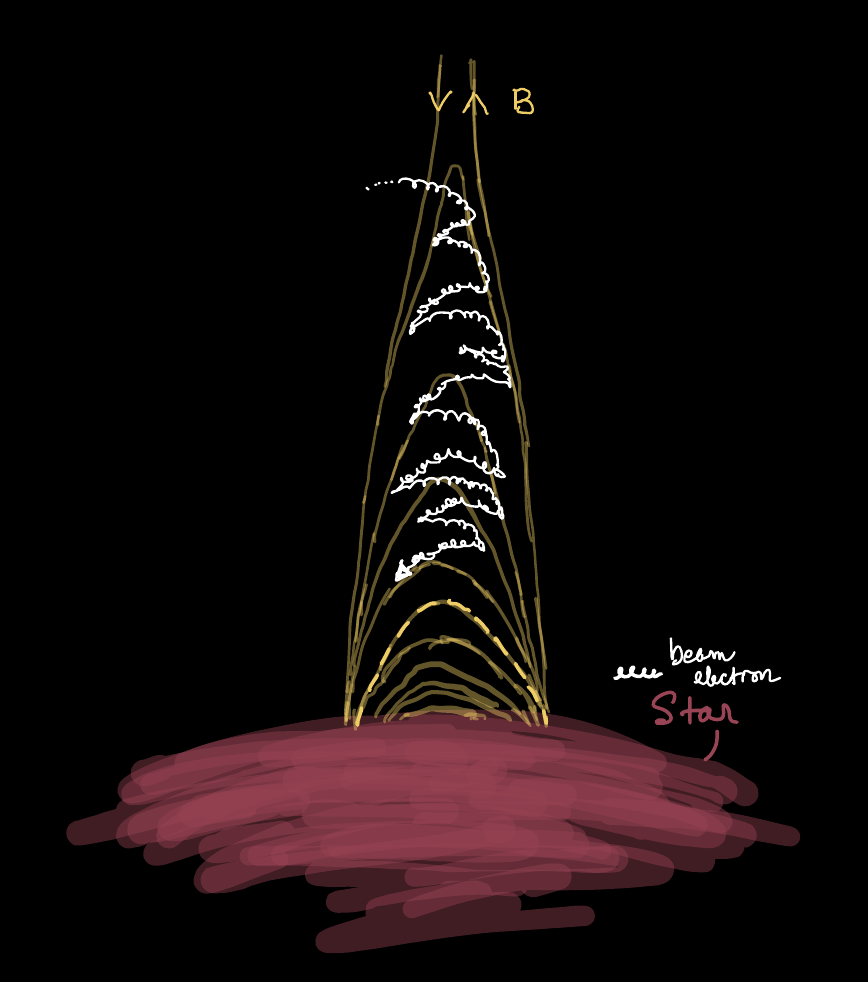}{0.4\textwidth}{(a)}
         \fig{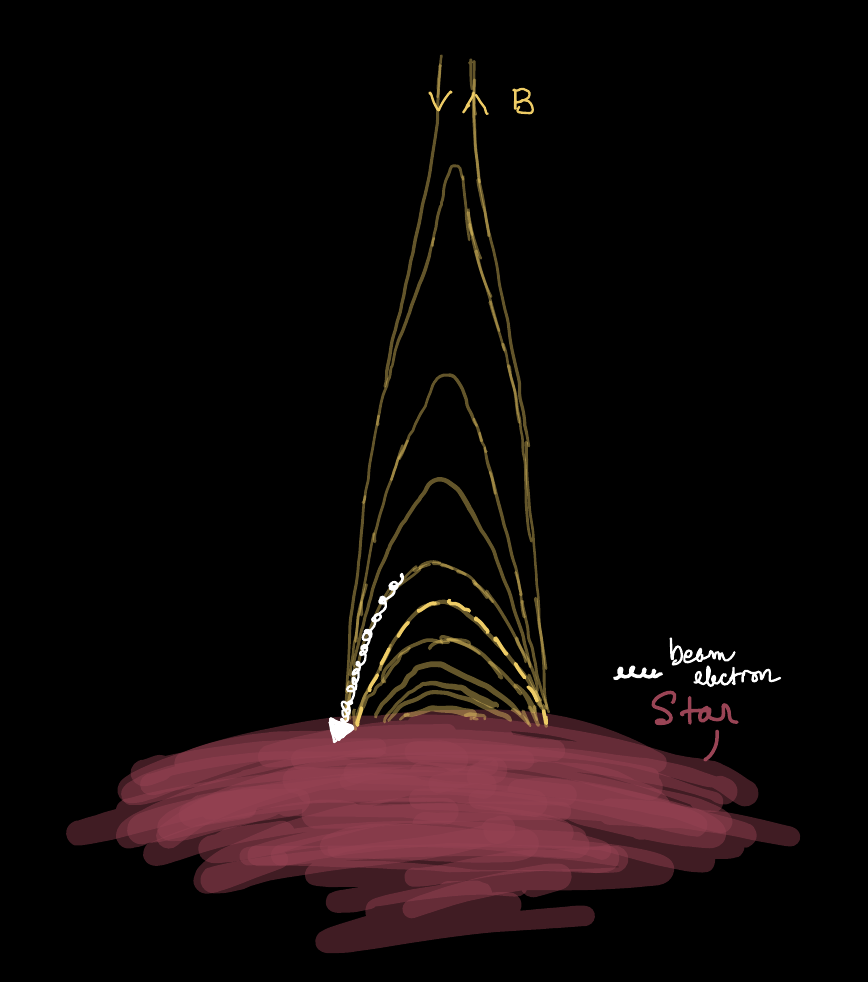}{0.4\textwidth}{(b)}
         }
\caption{ A hypothetical, two-step modular approach to modeling the time-dependent treatments of the K12 beam-plasma evolution in the corona and the radiative-hydrodynamic response of the lower atmosphere of a loop within a flare arcade. \textbf{ (a)} Illustrated time evolution of a reconnected field line retracting downward toward the relaxed, previously reconnected loops (dashed loop and below).  A hypothetical nonthermal electron path in the (pre-accelerated) power-law beam experiences additional acceleration to $E > 100$ keV through nonlinear wave-wave interactions as it is trapped in this region.   The illustrated geometries and particle paths (neither to scale) are largely inspired by the models of \cite{Aschwanden2004} and \cite{Egedal2015}. \textbf{(b)} Radiative-hydrodynamic modeling of the evolution of a semi-circular loop with field-aligned flows and plane-parallel radiative transfer, in response to heating by the K12 beam and a return current electric field, including any magnetic field convergence in the low atmosphere.  Note that the \cite{Aschwanden2004} prescription for pulsed nonthermal injection actually occurs during the retraction phase (panel a), but current RHD modeling capabilities with nonthermal particles cannot yet include the retraction of the magnetic field;  a short-duration constant flux injection of a K12 beam is likely sufficient for many purposes. }
\label{fig:cartoon}
\end{figure}

We suggest a modular modeling framework that stitches the current time-dependent K12 and steady-state \citep[e.g.][]{Allred2020} treatments of transport and heating in future RHD models of M dwarf flares.  This is illustrated in a sketch in Figure \ref{fig:cartoon}.  In panel (a), a reconnected magnetic field line retracts, and a nonthermal electron in the beam bounces back and forth due to a trapping mechanism  \citep[e.g.,][]{Aschwanden2004, Li2014, Egedal2015, Fleishman2022}.  It is conceivable that trapped beam particles have sufficient time for ``wave$+$wave, beam$+$wave'' processes to redistribute energy and enhance the number of electrons at $E > 100$ keV in this region.   The K12 theory is non-magnetic, but our prescription for beam injection into the RADYN model loop (Section \ref{sec:setup}) uses the pulsed injection prescription of \cite{Aschwanden2004} and emulates coronal trapping and injection timescales.  The evolution of the loss cone angle in \cite{Aschwanden2004} also might provide a mechanism to preserve beam anisotropy, which is required for the beam-plasma instability \citep[e.g.,][]{Tsytovich1995, Thorne2017}.  In panel (b), the K12 beam escapes to the footpoints and experiences a return current electric field and any possible magnetic convergence in the chromosphere.  The K12 beam in Figure \ref{fig:beamdep}(a) is expected to be significantly modified, especially at the low-energy end, by the return current electric field as formulated within the capabilities of current RHD modeling \citep{Holman2012, Allred2020}.  The RHD response of the chromosphere evolves, and plasma fills the relaxed magnetic loop.   The scenario in panel Figure \ref{fig:cartoon}(b) is ostensibly consistent with the free-streaming distances of accelerated electrons inferred from energy-dependent time delays of hard X-rays in the solar flare case \citep{Aschwanden1995, Aschwanden1996Masuda, Aschwanden1996, Aschwanden1996Conf}.  Note that the initial particle acceleration process(es) and location(s) are not specified in this rough sketch.  Of course, this scenario only intends to qualitatively show how the timescales may roughly fit together and provide a practical framework for models using current  capabilities rather than establish a completely self-consistent theory for all spatial, temporal, and spectral scales (which, to our knowledge, is an effort beyond the capabilities of any current methods of calculation for a realistic active star atmosphere).

The RHD modeling of the solar flare chromospheric response to the K12 modifications provides an alternative hypothesis test to the standard procedure that has been adopted in models of IRIS flare spectra \citep[e.g.,][]{Kuridze2015, Rubio2016, Kowalski2017Mar29}.  Several model shortcomings have been revealed in the relative brightness of the red-wing asymmetry emission component of Fe II and the emission around the rest wavelength at high-time resolution \citep{Graham2020}.  Injected beam distributions with larger relative heating rates in the stationary flare layers -- due to the enhancement of $E \gtrsim 100$ keV electrons in the beam -- may abate these discrepancies to some degree.  \citet{Kowalski2022} describe how the broadening of the optically thin, high-order hydrogen lines near the Balmer limit diagnoses the heating in the deeper, stationary chromosphere flare layers.  We plan to further develop these diagnostics to test the predictions of the heating rates from the K12 beams in RHD models.
Other difficulties that have been recently encountered in solar flare electron beam modeling, such as reproducing large continuum-to-line ratios in umbral flare brightenings \citep{KCF15, Kowalski2019IRIS} and so-called Type II white-light flare phenomena \citep{Hiei1982, Prochazka2019}, could also be investigated with RHD models that use injected particle flux distributions from the K12 theory.

K12 discusses the role of their beam transport modifications in reducing the number of required electrons to produce hard X-ray footpoint sources.   \citet{Hannah2013} include several of the effects from K12 into models of hard X-ray spectra from RHESSI, but to our knowledge, the non-linear, three-wave effects have not been quantitatively addressed and incorporated into the OSPEX X-ray modeling software.  
Here, we argue that the  calculations of K12 allow beams with large energy fluxes between $10^{12}$ and $10^{13}$ \fnum\ to propagate to the chromosphere in M-dwarf flares, resulting in energy deposition profiles similar to models with large low-energy cutoffs $E_c \gtrsim 85$ keV.   The integrated energy flux of the ``wave+wave, beam+wave'' spectrum in Figure \ref{fig:beamdep}(a) from K12 is $7\times10^8$ erg cm$^{-2}$ s$^{-1}$; this spectrum must be scaled by a factor of $\approx 1.7 \times 10^4$ for its heating rate to match the maximum heating rate in the cF13-85-3 simulation at log$_{10}\ m\ \approx -2$ (Figure \ref{fig:beamdep}(c)).  For this scaled K12 beam, the injected beam density at $E > 10$ keV is $\approx 10^{10}$ cm$^{-3}$.  \citet{Osten2006} infer compact regions with large coronal electron densities as large as $10^{12} - 10^{13}$ cm$^{-3}$ from quiescent, X-ray spectra of dMe stars.  We thus expect that the assumption of $n_{\rm{beam}} / n_{\rm{background}} = 10^{-2}$ employed in K12 to be valid, at least in some stellar active regions.  The $E \gtrsim 100$ keV electrons in the K12 beams also reach the chromosphere without significant collisional loss in such extremely dense coronae.  For larger relative beam densities, the ``wave$+$wave, beam$+$wave'' energy transfer processes are more important (E.\ Kontar, priv.\ communication, 2022), which can be understood as the result of more ion-acoustic turbulence present to enhance wave$+$wave processes.  At very large beam density, fluid and PIC simulations \citep{Bera2015, Bera2020} of ultra-relativistic, monoenergetic beams  instigate additional acceleration through the plasma wakefield effects that have only recently been considered in the context of solar/stellar flares \citep{Tsiklauri2017}.

Several physical processes in the numerical solutions of K12 have been investigated with PIC simulations on very short timescales.   The simulations of \citet{Karlicky2012} use a monoenergetic beam with a large density, $n_{\rm{beam}} / n_{\rm{background}} = 1/8$, and show  that Langmuir waves diffuse in $k$-space and boost some electrons above their initial energy \citep[see also the follow-up work of][]{Tsiklauri2014}.  They find that these effects occur before the onset of the Buneman instability in 1D PIC simulations \citep{Lee2008} of a drifting Gaussian beam that extends to relativistic energies.  \citet{Lee2011} use a 3D PIC code to simulate a narrower, drifting Gaussian electron beam, and they describe the energy exchange between the Gaussian beam and electromagnetic waves.  The kinetic energy of the drift is partially converted to thermal energy, widening the velocity distribution component parallel to the slowed drift.  \citet{Karlicky2009} and \citet{Karlicky2020} investigate PIC simulations with  power laws and kappa distributions.  Such beam distributions may eventually help bridge PIC models to flare observations of the chromosphere because of the important roles of the $E \gtrsim 100$ keV electrons in heating the low atmosphere and powering optical flare emission.  However, all PIC simulations in the present context do not include collisions, which self-consistently drive the bump-on-tail instability and Langmuir waves in the K12 theory.

\cite{Kowalski2022Frontiers} developed a  two-component RHD model of the rise and peak phase of the AD Leo Great Flare \citep{HP91}, which could be extended to the full evolution of all data constraints \citep[see discussion in][]{Kowalski2016} of the energetic dMe flare in Figure \ref{fig:yzcmi_flare}.  In this modeling approach, the late rise, peak, and early fast-decay phases consist of a large filling factor of bright kernels with $E_c \ge 85$ keV and large beam fluxes ($10^{13}$ erg cm$^{-2}$ s$^{-1}$) that dominate the continuum flux spectra with the color temperatures of $T_{\rm{col}} \approx 9,000 - 11,000$ K.  Additionally, lower beam flux models that generate chromospheric condensations with a larger filling factors perhaps represent larger area ribbons (following the analogies used in \citealt{Kowalski2022Frontiers}) that vary in relative area coverage between the two luminous peaks in Figure \ref{fig:yzcmi_flare}, thus giving different color temperatures by $\Delta T_{\rm{col}} \approx 1000$ K.  In the gradual decay phase, \cite{Kowalski2022Frontiers} speculated that the number of new bright kernels decrease relative to new bright ribbon areas while X-rays backheat the surrounding upper chromosphere \citep{HF92}, thus producing a lower optical color temperature between $6000-7000$ K.  With the detailed color information in the YZ CMi flare, this proposed multi-component spatial development can be explored using a parameter search of an M dwarf model grid for the lower flux component. An alternative RHD model superposition was proposed in \cite{Osten2016} and \cite{Kowalski2017Broadening} by replicating cool color temperatures in the decay phases of other very large dMe flares.

\section{Conclusions} \label{sec:conclusions}
High electron beam energy fluxes have been recently utilized in RHD models of M dwarf flares.  These reproduce the optical depths in the optical and ultraviolet continuum that are consistent with spectral observations.  However, most theoretical considerations suggest that beams with large current densities undergo systematic energy loss as they propagate to the chromosphere \citep[e.g.,][]{Oord1990, Zharkova2006, Lee2008, Holman2012, Alaoui2017, Allred2020}.  Other treatments, such as the K12 theory, predict a series of time-dependent, energy redistribution processes among beam particles and background plasma waves. The K12 theory has been compared with PIC simulations (where approximations allow), it predicts electron flux spectra that differ significantly from standard power laws inferred from collisional thick target modeling of solar flare hard X-rays, and it has yielded predictions of nonthermal radiative signatures \citep{Hannah2013, Ratcliffe2014}.  We find that the modifications to beam transport in K12 that include nonlinear wave-wave (scattering and decay) interactions and energy transfer between plasma waves and the beam produce a heating-rate maximum over the deep chromospheric column mass where the optical and NUV continuum radiation becomes optically thick if the temperatures increase enough.  Our predicted K12 heating rate in the deep chromosphere is  remarkably similar to the heating-rate distribution from an injected electron beam with a hard power-law distribution, $\delta = 3$, and a large low-energy cutoff, $E_c = 85$ keV.  These similarities owe to the large relative number of electrons with $E \gtrsim 100$ keV in both injected beam distributions.  Thus, we identify a tantalizing connection between semi-empirical RHD modeling approach with large, low-energy cutoff electron beams \citep{Kowalski2017Broadening} and the fundamental equations that describe the coupled time evolution of coronal plasma waves and the beam in the presence of Coulomb collisions.

The main purpose of this paper is to propose a physical explanation for high-energy stellar flare electron beams,
 which will comprise a large grid of publicly available RHD models.   We  demonstrated that one of the high-energy electron beam models, mF13-85-3, in this grid provides insight into the origin of the hot color temperatures that are inferred in optical narrowband continuum photometry and spectra in the impulsive phase of M dwarf flares.  The discrepancy between the values of $T_{\rm{gas}} > 12,000$ K over which the emergent continuum intensity forms and the color temperature, $T_{\rm{col}} < 11,000$ K of the emergent optical spectrum is indicative of a heating source in the low chromosphere that is more energetic than previously thought (e.g., $T \approx 9000$ K blackbody or optically thin hydrogen recombination).  This agrees with the conclusion in \citet{Kowalski2022Frontiers} that follows from their multi-component, emission line and continuum fitting to a well-studied M dwarf superflare.  Our chromospheric modeling approach bridges the coronal transport theory of K12 and suggests that the origin of this energy source is a large enhancement of power-law electrons at $E \gtrsim 100$ keV.

\section*{Acknowledgments}

AFK thanks an anonymous referee for interesting discussions and a critical review, which greatly improved the presentation of ideas in the paper.
   AFK gratefully acknowledges Eduard Kontar for helpful explanations and stimulating discussions about the results from \citet{Kontar2012}.  AFK acknowledges helpful discussions about the K12 results with Lyndsay Fletcher, Hugh Hudson, and Paulo Sim$\tilde{\rm{o}}$es.  AFK thanks Han Uitenbroek for discussions about magnetic field convergence properties in the chromosphere. AFK thanks Joel C. Allred and Mats Carlsson for developments to and assistance with the \texttt{RADYN} code.  AFK thanks Joel C. Allred for helpful dicussions about the Fokker-Planck formalism and magnetic mirror effects, and Marian Karlicky for a discussion about PIC simulations.  AFK thanks Rachel Osten for a reading of the manuscript.  AFK thanks Gregory Fleishman for discussions about the 2017 Sep 10 solar flare.  AFK thanks Jim Drake for helpful explanations about PIC simulation results pertaining to double layers.  AFK acknowledges funding support from NSF Award 1916511, NASA ADAP 80NSSC21K0632, and NASA grant 20-ECIP20\_2-0033.

\bibliography{final}{}

\begin{thebibliography}{}
\expandafter\ifx\csname natexlab\endcsname\relax\def\natexlab#1{#1}\fi
\providecommand{\url}[1]{\href{#1}{#1}}
\providecommand{\dodoi}[1]{doi:~\href{http://doi.org/#1}{\nolinkurl{#1}}}
\providecommand{\doeprint}[1]{\href{http://ascl.net/#1}{\nolinkurl{http://ascl.net/#1}}}
\providecommand{\doarXiv}[1]{\href{https://arxiv.org/abs/#1}{\nolinkurl{https://arxiv.org/abs/#1}}}

\bibitem[{{Abrevaya} {et~al.}(2020){Abrevaya}, {Leitzinger}, {Oppezzo},
  {Odert}, {Patel}, {Luna}, {Forte Giacobone}, \& {Hanslmeier}}]{Abrevaya2020}
{Abrevaya}, X.~C., {Leitzinger}, M., {Oppezzo}, O.~J., {et~al.} 2020, \mnras,
  494, L69, \dodoi{10.1093/mnrasl/slaa037}

\bibitem[{{Alaoui} \& {Holman}(2017)}]{Alaoui2017}
{Alaoui}, M., \& {Holman}, G.~D. 2017, \apj, 851, 78,
  \dodoi{10.3847/1538-4357/aa98de}

\bibitem[{{Allred} {et~al.}(2020){Allred}, {Alaoui}, {Kowalski}, \&
  {Kerr}}]{Allred2020}
{Allred}, J.~C., {Alaoui}, M., {Kowalski}, A.~F., \& {Kerr}, G.~S. 2020, \apj,
  902, 16, \dodoi{10.3847/1538-4357/abb239}

\bibitem[{{Allred} {et~al.}(2005){Allred}, {Hawley}, {Abbett}, \&
  {Carlsson}}]{Allred2005}
{Allred}, J.~C., {Hawley}, S.~L., {Abbett}, W.~P., \& {Carlsson}, M. 2005,
  \apj, 630, 573, \dodoi{10.1086/431751}

\bibitem[{{Allred} {et~al.}(2006){Allred}, {Hawley}, {Abbett}, \&
  {Carlsson}}]{Allred2006}
---. 2006, \apj, 644, 484, \dodoi{10.1086/503314}

\bibitem[{{Allred} {et~al.}(2015){Allred}, {Kowalski}, \&
  {Carlsson}}]{Allred2015}
{Allred}, J.~C., {Kowalski}, A.~F., \& {Carlsson}, M. 2015, \apj, 809, 104,
  \dodoi{10.1088/0004-637X/809/1/104}

\bibitem[{{Arber} {et~al.}(2015){Arber}, {Bennett}, {Brady},
  {Lawrence-Douglas}, {Ramsay}, {Sircombe}, {Gillies}, {Evans}, {Schmitz},
  {Bell}, \& {Ridgers}}]{Arber2015}
{Arber}, T.~D., {Bennett}, K., {Brady}, C.~S., {et~al.} 2015, Plasma Physics
  and Controlled Fusion, 57, 113001, \dodoi{10.1088/0741-3335/57/11/113001}

\bibitem[{{Aschwanden}(1996)}]{Aschwanden1996Conf}
{Aschwanden}, M.~J. 1996, in American Institute of Physics Conference Series,
  Vol. 374, High energy solar Physics, ed. R.~{Ramaty}, N.~{Mandzhavidze}, \&
  X.-M. {Hua}, 300--310, \dodoi{10.1063/1.50965}

\bibitem[{{Aschwanden}(2004)}]{Aschwanden2004}
{Aschwanden}, M.~J. 2004, \apj, 608, 554, \dodoi{10.1086/392494}

\bibitem[{{Aschwanden} {et~al.}(1996{\natexlab{a}}){Aschwanden}, {Hudson},
  {Kosugi}, \& {Schwartz}}]{Aschwanden1996Masuda}
{Aschwanden}, M.~J., {Hudson}, H., {Kosugi}, T., \& {Schwartz}, R.~A.
  1996{\natexlab{a}}, \apj, 464, 985, \dodoi{10.1086/177386}

\bibitem[{{Aschwanden} {et~al.}(1996{\natexlab{b}}){Aschwanden}, {Kosugi},
  {Hudson}, {Wills}, \& {Schwartz}}]{Aschwanden1996}
{Aschwanden}, M.~J., {Kosugi}, T., {Hudson}, H.~S., {Wills}, M.~J., \&
  {Schwartz}, R.~A. 1996{\natexlab{b}}, \apj, 470, 1198, \dodoi{10.1086/177943}

\bibitem[{{Aschwanden} {et~al.}(1995){Aschwanden}, {Schwartz}, \&
  {Alt}}]{Aschwanden1995}
{Aschwanden}, M.~J., {Schwartz}, R.~A., \& {Alt}, D.~M. 1995, \apj, 447, 923,
  \dodoi{10.1086/175930}

\bibitem[{{Battaglia} {et~al.}(2012){Battaglia}, {Kontar}, {Fletcher}, \&
  {MacKinnon}}]{Battaglia2012}
{Battaglia}, M., {Kontar}, E.~P., {Fletcher}, L., \& {MacKinnon}, A.~L. 2012,
  \apj, 752, 4, \dodoi{10.1088/0004-637X/752/1/4}

\bibitem[{{Ben{\'a}{\v{c}}ek} \& {Karlick{\'y}}(2020)}]{Karlicky2020}
{Ben{\'a}{\v{c}}ek}, J., \& {Karlick{\'y}}, M. 2020, \apj, 896, 9,
  \dodoi{10.3847/1538-4357/ab89a5}

\bibitem[{{Bera} {et~al.}(2020){Bera}, {Mandal}, {Das}, \&
  {Sengupta}}]{Bera2020}
{Bera}, R.~K., {Mandal}, D., {Das}, A., \& {Sengupta}, S. 2020, AIP Advances,
  10, 025203, \dodoi{10.1063/1.5126210}

\bibitem[{{Bera} {et~al.}(2015){Bera}, {Sengupta}, \& {Das}}]{Bera2015}
{Bera}, R.~K., {Sengupta}, S., \& {Das}, A. 2015, Physics of Plasmas, 22,
  073109, \dodoi{10.1063/1.4926816}

\bibitem[{{Brown}(1971)}]{Brown1971}
{Brown}, J.~C. 1971, \solphys, 18, 489, \dodoi{10.1007/BF00149070}

\bibitem[{{Brown} \& {Melrose}(1977)}]{Brown1977}
{Brown}, J.~C., \& {Melrose}, D.~B. 1977, \solphys, 52, 117,
  \dodoi{10.1007/BF00935795}

\bibitem[{{Brown} {et~al.}(2009){Brown}, {Turkmani}, {Kontar}, {MacKinnon}, \&
  {Vlahos}}]{Brown2009}
{Brown}, J.~C., {Turkmani}, R., {Kontar}, E.~P., {MacKinnon}, A.~L., \&
  {Vlahos}, L. 2009, \aap, 508, 993, \dodoi{10.1051/0004-6361/200913145}

\bibitem[{{Carlsson} \& {Stein}(1992)}]{Carlsson1992B}
{Carlsson}, M., \& {Stein}, R.~F. 1992, \apjl, 397, L59, \dodoi{10.1086/186544}

\bibitem[{{Carlsson} \& {Stein}(1995)}]{Carlsson1995}
---. 1995, \apjl, 440, L29, \dodoi{10.1086/187753}

\bibitem[{{Carlsson} \& {Stein}(1997)}]{Carlsson1997}
---. 1997, \apj, 481, 500

\bibitem[{{Carlsson} \& {Stein}(2002)}]{Carlsson2002}
---. 2002, \apj, 572, 626, \dodoi{10.1086/340293}

\bibitem[{{Chen} {et~al.}(2021){Chen}, {Zhan}, {Youngblood}, {Wolf},
  {Feinstein}, \& {Horton}}]{Chen2021}
{Chen}, H., {Zhan}, Z., {Youngblood}, A., {et~al.} 2021, Nature Astronomy, 5,
  298, \dodoi{10.1038/s41550-020-01264-1}

\bibitem[{{Cheng} {et~al.}(2010){Cheng}, {Ding}, \& {Carlsson}}]{Cheng2010}
{Cheng}, J.~X., {Ding}, M.~D., \& {Carlsson}, M. 2010, \apj, 711, 185,
  \dodoi{10.1088/0004-637X/711/1/185}

\bibitem[{{Christian} {et~al.}(2003){Christian}, {Mathioudakis},
  {Jevremovi{\'c}}, {Dupuis}, {Vennes}, \& {Kawka}}]{Christian2003}
{Christian}, D.~J., {Mathioudakis}, M., {Jevremovi{\'c}}, D., {et~al.} 2003,
  \apjl, 593, L105, \dodoi{10.1086/378217}

\bibitem[{{Cram} \& {Woods}(1982)}]{Cram1982}
{Cram}, L.~E., \& {Woods}, D.~T. 1982, \apj, 257, 269, \dodoi{10.1086/159985}

\bibitem[{{Dennis} {et~al.}(2022){Dennis}, {Shih}, {Hurford}, \&
  {Saint-Hilaire}}]{Dennis2022}
{Dennis}, B.~R., {Shih}, A.~Y., {Hurford}, G.~J., \& {Saint-Hilaire}, P. 2022,
  arXiv e-prints, arXiv:2206.00741.
\newblock \doarXiv{2206.00741}

\bibitem[{{Dennis} \& {Tolbert}(2019)}]{Dennis2019}
{Dennis}, B.~R., \& {Tolbert}, A.~K. 2019, \apj, 887, 131,
  \dodoi{10.3847/1538-4357/ab4f81}

\bibitem[{{Dennis} \& {Zarro}(1993)}]{Dennis1993}
{Dennis}, B.~R., \& {Zarro}, D.~M. 1993, \solphys, 146, 177,
  \dodoi{10.1007/BF00662178}

\bibitem[{{Dhillon} {et~al.}(2007){Dhillon}, {Marsh}, {Stevenson}, {Atkinson},
  {Kerry}, {Peacocke}, {Vick}, {Beard}, {Ives}, {Lunney}, {McLay}, {Tierney},
  {Kelly}, {Littlefair}, {Nicholson}, {Pashley}, {Harlaftis}, \&
  {O'Brien}}]{Dhillon2007}
{Dhillon}, V.~S., {Marsh}, T.~R., {Stevenson}, M.~J., {et~al.} 2007, \mnras,
  378, 825, \dodoi{10.1111/j.1365-2966.2007.11881.x}

\bibitem[{{Dominique} {et~al.}(2018){Dominique}, {Zhukov}, {Heinzel},
  {Dammasch}, {Wauters}, {Dolla}, {Shestov}, {Kretzschmar}, {Machol},
  {Lapenta}, \& {Schmutz}}]{Dominique2018}
{Dominique}, M., {Zhukov}, A.~N., {Heinzel}, P., {et~al.} 2018, \apjl, 867,
  L24, \dodoi{10.3847/2041-8213/aaeace}

\bibitem[{{Dorfi} \& {Drury}(1987)}]{Dorfi1987}
{Dorfi}, E.~A., \& {Drury}, L.~O. 1987, Journal of Computational Physics, 69,
  175, \dodoi{10.1016/0021-9991(87)90161-6}

\bibitem[{{Egedal} {et~al.}(2015){Egedal}, {Daughton}, {Le}, \&
  {Borg}}]{Egedal2015}
{Egedal}, J., {Daughton}, W., {Le}, A., \& {Borg}, A.~L. 2015, Physics of
  Plasmas, 22, 101208, \dodoi{10.1063/1.4933055}

\bibitem[{{Emslie}(1978)}]{Emslie1978}
{Emslie}, A.~G. 1978, \apj, 224, 241, \dodoi{10.1086/156371}

\bibitem[{{Fleishman} {et~al.}(2022){Fleishman}, {Nita}, {Chen}, {Yu}, \&
  {Gary}}]{Fleishman2022}
{Fleishman}, G.~D., {Nita}, G.~M., {Chen}, B., {Yu}, S., \& {Gary}, D.~E. 2022,
  \nat, 606, 674, \dodoi{10.1038/s41586-022-04728-8}

\bibitem[{{Fuhrmeister} {et~al.}(2008){Fuhrmeister}, {Liefke}, {Schmitt}, \&
  {Reiners}}]{Fuhrmeister2008}
{Fuhrmeister}, B., {Liefke}, C., {Schmitt}, J.~H.~M.~M., \& {Reiners}, A. 2008,
  \aap, 487, 293, \dodoi{10.1051/0004-6361:200809379}

\bibitem[{{Fuhrmeister} {et~al.}(2010){Fuhrmeister}, {Schmitt}, \&
  {Hauschildt}}]{Fuhrmeister2010}
{Fuhrmeister}, B., {Schmitt}, J.~H.~M.~M., \& {Hauschildt}, P.~H. 2010, \aap,
  511, A83, \dodoi{10.1051/0004-6361/200810224}

\bibitem[{{Graham} {et~al.}(2020){Graham}, {Cauzzi}, {Zangrilli}, {Kowalski},
  {Sim{\~o}es}, \& {Allred}}]{Graham2020}
{Graham}, D.~R., {Cauzzi}, G., {Zangrilli}, L., {et~al.} 2020, \apj, 895, 6,
  \dodoi{10.3847/1538-4357/ab88ad}

\bibitem[{{Guedel} {et~al.}(1996){Guedel}, {Benz}, {Schmitt}, \&
  {Skinner}}]{Gudel1996}
{Guedel}, M., {Benz}, A.~O., {Schmitt}, J.~H.~M.~M., \& {Skinner}, S.~L. 1996,
  \apj, 471, 1002, \dodoi{10.1086/178027}

\bibitem[{{G{\"u}nther} {et~al.}(2020){G{\"u}nther}, {Zhan}, {Seager},
  {Rimmer}, {Ranjan}, {Stassun}, {Oelkers}, {Daylan}, {Newton}, {Kristiansen},
  {Olah}, {Gillen}, {Rappaport}, {Ricker}, {Vanderspek}, {Latham}, {Winn},
  {Jenkins}, {Glidden}, {Fausnaugh}, {Levine}, {Dittmann}, {Quinn},
  {Krishnamurthy}, \& {Ting}}]{Gunther2020}
{G{\"u}nther}, M.~N., {Zhan}, Z., {Seager}, S., {et~al.} 2020, \aj, 159, 60,
  \dodoi{10.3847/1538-3881/ab5d3a}

\bibitem[{{Hamilton} \& {Petrosian}(1987)}]{Hamilton1987}
{Hamilton}, R.~J., \& {Petrosian}, V. 1987, \apj, 321, 721,
  \dodoi{10.1086/165665}

\bibitem[{{Hannah} {et~al.}(2013){Hannah}, {Kontar}, \& {Reid}}]{Hannah2013}
{Hannah}, I.~G., {Kontar}, E.~P., \& {Reid}, H.~A.~S. 2013, \aap, 550, A51,
  \dodoi{10.1051/0004-6361/201220462}

\bibitem[{{Hannah} {et~al.}(2009){Hannah}, {Kontar}, \& {Sirenko}}]{Hannah2009}
{Hannah}, I.~G., {Kontar}, E.~P., \& {Sirenko}, O.~K. 2009, \apjl, 707, L45,
  \dodoi{10.1088/0004-637X/707/1/L45}

\bibitem[{{Hawley} {et~al.}(2014){Hawley}, {Davenport}, {Kowalski},
  {Wisniewski}, {Hebb}, {Deitrick}, \& {Hilton}}]{Hawley2014}
{Hawley}, S.~L., {Davenport}, J.~R.~A., {Kowalski}, A.~F., {et~al.} 2014, \apj,
  797, 121, \dodoi{10.1088/0004-637X/797/2/121}

\bibitem[{{Hawley} \& {Fisher}(1992)}]{HF92}
{Hawley}, S.~L., \& {Fisher}, G.~H. 1992, \apjs, 78, 565,
  \dodoi{10.1086/191640}

\bibitem[{{Hawley} \& {Fisher}(1994)}]{HF94}
---. 1994, \apj, 426, 387, \dodoi{10.1086/174075}

\bibitem[{{Hawley} \& {Pettersen}(1991)}]{HP91}
{Hawley}, S.~L., \& {Pettersen}, B.~R. 1991, \apj, 378, 725,
  \dodoi{10.1086/170474}

\bibitem[{{Hawley} {et~al.}(1995){Hawley}, {Fisher}, {Simon}, {Cully},
  {Deustua}, {Jablonski}, {Johns-Krull}, {Pettersen}, {Smith}, {Spiesman}, \&
  {Valenti}}]{Hawley1995}
{Hawley}, S.~L., {Fisher}, G.~H., {Simon}, T., {et~al.} 1995, \apj, 453, 464,
  \dodoi{10.1086/176408}

\bibitem[{{Heinzel} \& {Kleint}(2014)}]{Heinzel2014}
{Heinzel}, P., \& {Kleint}, L. 2014, \apjl, 794, L23,
  \dodoi{10.1088/2041-8205/794/2/L23}

\bibitem[{{Hiei}(1982)}]{Hiei1982}
{Hiei}, E. 1982, \solphys, 80, 113, \dodoi{10.1007/BF00153427}

\bibitem[{{Holman}(2012)}]{Holman2012}
{Holman}, G.~D. 2012, \apj, 745, 52, \dodoi{10.1088/0004-637X/745/1/52}

\bibitem[{{Holman} {et~al.}(2003){Holman}, {Sui}, {Schwartz}, \&
  {Emslie}}]{Holman2003}
{Holman}, G.~D., {Sui}, L., {Schwartz}, R.~A., \& {Emslie}, A.~G. 2003, \apjl,
  595, L97, \dodoi{10.1086/378488}

\bibitem[{{Howard} {et~al.}(2018){Howard}, {Tilley}, {Corbett}, {Youngblood},
  {Loyd}, {Ratzloff}, {Law}, {Fors}, {del Ser}, {Shkolnik}, {Ziegler}, {Goeke},
  {Pietraallo}, \& {Haislip}}]{Howard2018}
{Howard}, W.~S., {Tilley}, M.~A., {Corbett}, H., {et~al.} 2018, \apjl, 860,
  L30, \dodoi{10.3847/2041-8213/aacaf3}

\bibitem[{{Howard} {et~al.}(2020){Howard}, {Corbett}, {Law}, {Ratzloff},
  {Galliher}, {Glazier}, {Gonzalez}, {Vasquez Soto}, {Fors}, {del Ser}, \&
  {Haislip}}]{Howard2020}
{Howard}, W.~S., {Corbett}, H., {Law}, N.~M., {et~al.} 2020, \apj, 902, 115,
  \dodoi{10.3847/1538-4357/abb5b4}

\bibitem[{{Ireland} {et~al.}(2013){Ireland}, {Tolbert}, {Schwartz}, {Holman},
  \& {Dennis}}]{Ireland2013}
{Ireland}, J., {Tolbert}, A.~K., {Schwartz}, R.~A., {Holman}, G.~D., \&
  {Dennis}, B.~R. 2013, \apj, 769, 89, \dodoi{10.1088/0004-637X/769/2/89}

\bibitem[{{Karlick{\'y}} \& {Ka{\v{s}}parov{\'a}}(2009)}]{Karlicky2009}
{Karlick{\'y}}, M., \& {Ka{\v{s}}parov{\'a}}, J. 2009, \aap, 506, 1437,
  \dodoi{10.1051/0004-6361/200912616}

\bibitem[{{Karlick{\'y}} \& {Kontar}(2012)}]{Karlicky2012}
{Karlick{\'y}}, M., \& {Kontar}, E.~P. 2012, \aap, 544, A148,
  \dodoi{10.1051/0004-6361/201219400}

\bibitem[{{Kawate} {et~al.}(2012){Kawate}, {Nishizuka}, {Oi}, {Ohyama}, \&
  {Nakajima}}]{Kawate2012}
{Kawate}, T., {Nishizuka}, N., {Oi}, A., {Ohyama}, M., \& {Nakajima}, H. 2012,
  \apj, 747, 131, \dodoi{10.1088/0004-637X/747/2/131}

\bibitem[{{Kleint} {et~al.}(2016){Kleint}, {Heinzel}, {Judge}, \&
  {Krucker}}]{Kleint2016}
{Kleint}, L., {Heinzel}, P., {Judge}, P., \& {Krucker}, S. 2016, \apj, 816, 88,
  \dodoi{10.3847/0004-637X/816/2/88}

\bibitem[{{Kontar}(2001)}]{Kontar2001}
{Kontar}, E.~P. 2001, Computer Physics Communications, 138, 222,
  \dodoi{10.1016/S0010-4655(01)00214-4}

\bibitem[{{Kontar} {et~al.}(2008){Kontar}, {Hannah}, \&
  {MacKinnon}}]{Kontar2008}
{Kontar}, E.~P., {Hannah}, I.~G., \& {MacKinnon}, A.~L. 2008, \aap, 489, L57,
  \dodoi{10.1051/0004-6361:200810719}

\bibitem[{{Kontar} \& {P{\'e}cseli}(2002)}]{Kontar2002}
{Kontar}, E.~P., \& {P{\'e}cseli}, H.~L. 2002, \pre, 65, 066408,
  \dodoi{10.1103/PhysRevE.65.066408}

\bibitem[{{Kontar} {et~al.}(2012){Kontar}, {Ratcliffe}, \& {Bian}}]{Kontar2012}
{Kontar}, E.~P., {Ratcliffe}, H., \& {Bian}, N.~H. 2012, \aap, 539, A43,
  \dodoi{10.1051/0004-6361/201118216}

\bibitem[{{Kontar} {et~al.}(2011){Kontar}, {Brown}, {Emslie}, {Hajdas},
  {Holman}, {Hurford}, {Ka{\v{s}}parov{\'a}}, {Mallik}, {Massone}, {McConnell},
  {Piana}, {Prato}, {Schmahl}, \& {Suarez-Garcia}}]{Kontar2011}
{Kontar}, E.~P., {Brown}, J.~C., {Emslie}, A.~G., {et~al.} 2011, \ssr, 159,
  301, \dodoi{10.1007/s11214-011-9804-x}

\bibitem[{{Kowalski}(2022)}]{Kowalski2022Frontiers}
{Kowalski}, A.~F. 2022, Frontiers in Astronomy and Space Sciences, 9, 1034458,
  \dodoi{10.3389/fspas.2022.1034458}

\bibitem[{{Kowalski} {et~al.}(2022){Kowalski}, {Allred}, {Carlsson}, {Kerr},
  {Tremblay}, {Namekata}, {Kuridze}, \& {Uitenbroek}}]{Kowalski2022}
{Kowalski}, A.~F., {Allred}, J.~C., {Carlsson}, M., {et~al.} 2022, \apj, 928,
  190, \dodoi{10.3847/1538-4357/ac5174}

\bibitem[{{Kowalski} {et~al.}(2017{\natexlab{a}}){Kowalski}, {Allred}, {Daw},
  {Cauzzi}, \& {Carlsson}}]{Kowalski2017Mar29}
{Kowalski}, A.~F., {Allred}, J.~C., {Daw}, A., {Cauzzi}, G., \& {Carlsson}, M.
  2017{\natexlab{a}}, \apj, 836, 12, \dodoi{10.3847/1538-4357/836/1/12}

\bibitem[{{Kowalski} {et~al.}(2019{\natexlab{a}}){Kowalski}, {Butler}, {Daw},
  {Fletcher}, {Allred}, {De Pontieu}, {Kerr}, \& {Cauzzi}}]{Kowalski2019IRIS}
{Kowalski}, A.~F., {Butler}, E., {Daw}, A.~N., {et~al.} 2019{\natexlab{a}},
  \apj, 878, 135, \dodoi{10.3847/1538-4357/ab1f8b}

\bibitem[{{Kowalski} {et~al.}(2015{\natexlab{a}}){Kowalski}, {Cauzzi}, \&
  {Fletcher}}]{KCF15}
{Kowalski}, A.~F., {Cauzzi}, G., \& {Fletcher}, L. 2015{\natexlab{a}}, \apj,
  798, 107, \dodoi{10.1088/0004-637X/798/2/107}

\bibitem[{{Kowalski} {et~al.}(2015{\natexlab{b}}){Kowalski}, {Hawley},
  {Carlsson}, {Allred}, {Uitenbroek}, {Osten}, \& {Holman}}]{Kowalski2015}
{Kowalski}, A.~F., {Hawley}, S.~L., {Carlsson}, M., {et~al.}
  2015{\natexlab{b}}, \solphys, 290, 3487, \dodoi{10.1007/s11207-015-0708-x}

\bibitem[{{Kowalski} {et~al.}(2011){Kowalski}, {Hawley}, {Holtzman},
  {Wisniewski}, \& {Hilton}}]{Kowalski2011}
{Kowalski}, A.~F., {Hawley}, S.~L., {Holtzman}, J.~A., {Wisniewski}, J.~P., \&
  {Hilton}, E.~J. 2011, in Physics of Sun and Star Spots, ed. D.~{Prasad
  Choudhary} \& K.~G. {Strassmeier}, Vol. 273, 261--264,
  \dodoi{10.1017/S1743921311015341}

\bibitem[{{Kowalski} {et~al.}(2013){Kowalski}, {Hawley}, {Wisniewski}, {Osten},
  {Hilton}, {Holtzman}, {Schmidt}, \& {Davenport}}]{Kowalski2013}
{Kowalski}, A.~F., {Hawley}, S.~L., {Wisniewski}, J.~P., {et~al.} 2013, \apjs,
  207, 15, \dodoi{10.1088/0067-0049/207/1/15}

\bibitem[{{Kowalski} {et~al.}(2018){Kowalski}, {Mathioudakis}, \&
  {Hawley}}]{Kowalski2018}
{Kowalski}, A.~F., {Mathioudakis}, M., \& {Hawley}, S.~L. 2018, in 20th
  Cambridge Workshop on Cool Stars, Stellar Systems and the Sun, Cambridge
  Workshop on Cool Stars, Stellar Systems, and the Sun, 42,
  \dodoi{10.5281/zenodo.1463140}

\bibitem[{{Kowalski} {et~al.}(2016){Kowalski}, {Mathioudakis}, {Hawley},
  {Wisniewski}, {Dhillon}, {Marsh}, {Hilton}, \& {Brown}}]{Kowalski2016}
{Kowalski}, A.~F., {Mathioudakis}, M., {Hawley}, S.~L., {et~al.} 2016, \apj,
  820, 95, \dodoi{10.3847/0004-637X/820/2/95}

\bibitem[{{Kowalski} {et~al.}(2017{\natexlab{b}}){Kowalski}, {Allred},
  {Uitenbroek}, {Tremblay}, {Brown}, {Carlsson}, {Osten}, {Wisniewski}, \&
  {Hawley}}]{Kowalski2017Broadening}
{Kowalski}, A.~F., {Allred}, J.~C., {Uitenbroek}, H., {et~al.}
  2017{\natexlab{b}}, \apj, 837, 125, \dodoi{10.3847/1538-4357/aa603e}

\bibitem[{{Kowalski} {et~al.}(2019{\natexlab{b}}){Kowalski}, {Wisniewski},
  {Hawley}, {Osten}, {Brown}, {Fari{\~n}a}, {Valenti}, {Brown}, {Xilouris},
  {Schmidt}, \& {Johns-Krull}}]{Kowalski2019HST}
{Kowalski}, A.~F., {Wisniewski}, J.~P., {Hawley}, S.~L., {et~al.}
  2019{\natexlab{b}}, \apj, 871, 167, \dodoi{10.3847/1538-4357/aaf058}

\bibitem[{{Krucker} {et~al.}(2011){Krucker}, {Hudson}, {Jeffrey}, {Battaglia},
  {Kontar}, {Benz}, {Csillaghy}, \& {Lin}}]{Krucker2011}
{Krucker}, S., {Hudson}, H.~S., {Jeffrey}, N.~L.~S., {et~al.} 2011, \apj, 739,
  96, \dodoi{10.1088/0004-637X/739/2/96}

\bibitem[{{Kundu} {et~al.}(2009){Kundu}, {Grechnev}, {White}, {Schmahl},
  {Meshalkina}, \& {Kashapova}}]{Kundu2009}
{Kundu}, M.~R., {Grechnev}, V.~V., {White}, S.~M., {et~al.} 2009, \solphys,
  260, 135, \dodoi{10.1007/s11207-009-9437-3}

\bibitem[{{Kuridze} {et~al.}(2015){Kuridze}, {Mathioudakis}, {Sim{\~o}es},
  {Rouppe van der Voort}, {Carlsson}, {Jafarzadeh}, {Allred}, {Kowalski},
  {Kennedy}, {Fletcher}, {Graham}, \& {Keenan}}]{Kuridze2015}
{Kuridze}, D., {Mathioudakis}, M., {Sim{\~o}es}, P.~J.~A., {et~al.} 2015, \apj,
  813, 125, \dodoi{10.1088/0004-637X/813/2/125}

\bibitem[{{Leach} \& {Petrosian}(1981)}]{Leach1981}
{Leach}, J., \& {Petrosian}, V. 1981, \apj, 251, 781, \dodoi{10.1086/159521}

\bibitem[{{Lee} \& {B{\"u}chner}(2011)}]{Lee2011}
{Lee}, K.~W., \& {B{\"u}chner}, J. 2011, \aap, 535, A61,
  \dodoi{10.1051/0004-6361/201117186}

\bibitem[{{Lee} {et~al.}(2008){Lee}, {B{\"u}chner}, \& {Elkina}}]{Lee2008}
{Lee}, K.~W., {B{\"u}chner}, J., \& {Elkina}, N. 2008, \aap, 478, 889,
  \dodoi{10.1051/0004-6361:20078419}

\bibitem[{{Li} {et~al.}(2012){Li}, {Drake}, \& {Swisdak}}]{Li2012}
{Li}, T.~C., {Drake}, J.~F., \& {Swisdak}, M. 2012, \apj, 757, 20,
  \dodoi{10.1088/0004-637X/757/1/20}

\bibitem[{{Li} {et~al.}(2014){Li}, {Drake}, \& {Swisdak}}]{Li2014}
---. 2014, \apj, 793, 7, \dodoi{10.1088/0004-637X/793/1/7}

\bibitem[{{Livshits} {et~al.}(1981){Livshits}, {Badalian}, {Kosovichev}, \&
  {Katsova}}]{Livshits1981}
{Livshits}, M.~A., {Badalian}, O.~G., {Kosovichev}, A.~G., \& {Katsova}, M.~M.
  1981, \solphys, 73, 269, \dodoi{10.1007/BF00151682}

\bibitem[{{Loyd} {et~al.}(2018){Loyd}, {France}, {Youngblood}, {Schneider},
  {Brown}, {Hu}, {Segura}, {Linsky}, {Redfield}, {Tian}, {Rugheimer}, {Miguel},
  \& {Froning}}]{Loyd2018}
{Loyd}, R.~O.~P., {France}, K., {Youngblood}, A., {et~al.} 2018, \apj, 867, 71,
  \dodoi{10.3847/1538-4357/aae2bd}

\bibitem[{{MacGregor} {et~al.}(2020){MacGregor}, {Osten}, \&
  {Hughes}}]{MacGregor2020}
{MacGregor}, A.~M., {Osten}, R.~A., \& {Hughes}, A.~M. 2020, \apj, 891, 80,
  \dodoi{10.3847/1538-4357/ab711d}

\bibitem[{{MacGregor} {et~al.}(2018){MacGregor}, {Weinberger}, {Wilner},
  {Kowalski}, \& {Cranmer}}]{MacGregor2018}
{MacGregor}, M.~A., {Weinberger}, A.~J., {Wilner}, D.~J., {Kowalski}, A.~F., \&
  {Cranmer}, S.~R. 2018, \apjl, 855, L2, \dodoi{10.3847/2041-8213/aaad6b}

\bibitem[{{MacGregor} {et~al.}(2021){MacGregor}, {Weinberger}, {Loyd},
  {Shkolnik}, {Barclay}, {Howard}, {Zic}, {Osten}, {Cranmer}, {Kowalski},
  {Lenc}, {Youngblood}, {Estes}, {Wilner}, {Forbrich}, {Hughes}, {Law},
  {Murphy}, {Boley}, \& {Matthews}}]{MacGregor2021}
{MacGregor}, M.~A., {Weinberger}, A.~J., {Loyd}, R.~O.~P., {et~al.} 2021,
  \apjl, 911, L25, \dodoi{10.3847/2041-8213/abf14c}

\bibitem[{{Maehara} {et~al.}(2021){Maehara}, {Notsu}, {Namekata}, {Honda},
  {Kowalski}, {Katoh}, {Ohshima}, {Iida}, {Oeda}, {Murata}, {Yamanaka},
  {Takagi}, {Sasada}, {Akitaya}, {Ikuta}, {Okamoto}, {Nogami}, \&
  {Shibata}}]{Maehara2021}
{Maehara}, H., {Notsu}, Y., {Namekata}, K., {et~al.} 2021, \pasj, 73, 44,
  \dodoi{10.1093/pasj/psaa098}

\bibitem[{{Mart{\'\i}nez Oliveros} {et~al.}(2012){Mart{\'\i}nez Oliveros},
  {Hudson}, {Hurford}, {Krucker}, {Lin}, {Lindsey}, {Couvidat}, {Schou}, \&
  {Thompson}}]{Martinez2012}
{Mart{\'\i}nez Oliveros}, J.-C., {Hudson}, H.~S., {Hurford}, G.~J., {et~al.}
  2012, \apjl, 753, L26, \dodoi{10.1088/2041-8205/753/2/L26}

\bibitem[{{McTiernan} \& {Petrosian}(1990)}]{McTiernan1990}
{McTiernan}, J.~M., \& {Petrosian}, V. 1990, \apj, 359, 524,
  \dodoi{10.1086/169084}

\bibitem[{{Milligan} {et~al.}(2014){Milligan}, {Kerr}, {Dennis}, {Hudson},
  {Fletcher}, {Allred}, {Chamberlin}, {Ireland}, {Mathioudakis}, \&
  {Keenan}}]{Milligan2014}
{Milligan}, R.~O., {Kerr}, G.~S., {Dennis}, B.~R., {et~al.} 2014, \apj, 793,
  70, \dodoi{10.1088/0004-637X/793/2/70}

\bibitem[{{Mochnacki} \& {Zirin}(1980)}]{Mochnacki1980}
{Mochnacki}, S.~W., \& {Zirin}, H. 1980, \apjl, 239, L27,
  \dodoi{10.1086/183285}

\bibitem[{{Namekata} {et~al.}(2020){Namekata}, {Maehara}, {Sasaki}, {Kawai},
  {Notsu}, {Kowalski}, {Allred}, {Iwakiri}, {Tsuboi}, {Murata}, {Niwano},
  {Shiraishi}, {Adachi}, {Iida}, {Oeda}, {Honda}, {Tozuka}, {Katoh}, {Onozato},
  {Okamoto}, {Isogai}, {Kimura}, {Kojiguchi}, {Wakamatsu}, {Tampo}, {Nogami},
  \& {Shibata}}]{Namekata2020}
{Namekata}, K., {Maehara}, H., {Sasaki}, R., {et~al.} 2020, \pasj,
  \dodoi{10.1093/pasj/psaa051}

\bibitem[{{Neidig}(1983)}]{Neidig1983}
{Neidig}, D.~F. 1983, \solphys, 85, 285, \dodoi{10.1007/BF00148655}

\bibitem[{{Neidig} {et~al.}(1993){Neidig}, {Kiplinger}, {Cohl}, \&
  {Wiborg}}]{Neidig1993}
{Neidig}, D.~F., {Kiplinger}, A.~L., {Cohl}, H.~S., \& {Wiborg}, P.~H. 1993,
  \apj, 406, 306, \dodoi{10.1086/172442}

\bibitem[{{Neupert}(1968)}]{Neupert1968}
{Neupert}, W.~M. 1968, \apjl, 153, L59, \dodoi{10.1086/180220}

\bibitem[{{Nishikawa} {et~al.}(2021){Nishikawa}, {Du{\c{t}}an}, {K{\"o}hn}, \&
  {Mizuno}}]{Nishikawa2021}
{Nishikawa}, K., {Du{\c{t}}an}, I., {K{\"o}hn}, C., \& {Mizuno}, Y. 2021,
  Living Reviews in Computational Astrophysics, 7, 1,
  \dodoi{10.1007/s41115-021-00012-0}

\bibitem[{{Osten} {et~al.}(2007){Osten}, {Drake}, {Tueller}, {Cummings},
  {Perri}, {Moretti}, \& {Covino}}]{Osten2007}
{Osten}, R.~A., {Drake}, S., {Tueller}, J., {et~al.} 2007, \apj, 654, 1052,
  \dodoi{10.1086/509252}

\bibitem[{{Osten} {et~al.}(2006){Osten}, {Hawley}, {Allred}, {Johns-Krull},
  {Brown}, \& {Harper}}]{Osten2006}
{Osten}, R.~A., {Hawley}, S.~L., {Allred}, J., {et~al.} 2006, \apj, 647, 1349,
  \dodoi{10.1086/504889}

\bibitem[{{Osten} {et~al.}(2016){Osten}, {Kowalski}, {Drake}, {Krimm}, {Page},
  {Gazeas}, {Kennea}, {Oates}, {Page}, {de Miguel}, {Nov{\'a}k}, {Apeltauer},
  \& {Gehrels}}]{Osten2016}
{Osten}, R.~A., {Kowalski}, A., {Drake}, S.~A., {et~al.} 2016, \apj, 832, 174,
  \dodoi{10.3847/0004-637X/832/2/174}

\bibitem[{{Pechhacker} \& {Tsiklauri}(2014)}]{Tsiklauri2014}
{Pechhacker}, R., \& {Tsiklauri}, D. 2014, Physics of Plasmas, 21, 012903,
  \dodoi{10.1063/1.4863494}

\bibitem[{{Potts} {et~al.}(2010){Potts}, {Hudson}, {Fletcher}, \&
  {Diver}}]{Potts2010}
{Potts}, H., {Hudson}, H., {Fletcher}, L., \& {Diver}, D. 2010, \apj, 722,
  1514, \dodoi{10.1088/0004-637X/722/2/1514}

\bibitem[{{Proch{\'a}zka} {et~al.}(2017){Proch{\'a}zka}, {Milligan}, {Allred},
  {Kowalski}, {Kotr{\v c}}, \& {Mathioudakis}}]{Ondrej1}
{Proch{\'a}zka}, O., {Milligan}, R.~O., {Allred}, J.~C., {et~al.} 2017, \apj,
  837, 46, \dodoi{10.3847/1538-4357/aa5da8}

\bibitem[{{Proch{\'a}zka} {et~al.}(2019){Proch{\'a}zka}, {Reid}, \&
  {Mathioudakis}}]{Prochazka2019}
{Proch{\'a}zka}, O., {Reid}, A., \& {Mathioudakis}, M. 2019, \apj, 882, 97,
  \dodoi{10.3847/1538-4357/ab35e1}

\bibitem[{{Ratcliffe} {et~al.}(2012){Ratcliffe}, {Bian}, \&
  {Kontar}}]{Ratcliffe2012}
{Ratcliffe}, H., {Bian}, N.~H., \& {Kontar}, E.~P. 2012, \apj, 761, 176,
  \dodoi{10.1088/0004-637X/761/2/176}

\bibitem[{{Ratcliffe} \& {Kontar}(2014)}]{Ratcliffe2014}
{Ratcliffe}, H., \& {Kontar}, E.~P. 2014, \aap, 562, A57,
  \dodoi{10.1051/0004-6361/201322263}

\bibitem[{{Raulin} {et~al.}(2004){Raulin}, {Makhmutov}, {Kaufmann}, {Pacini},
  {L{\"u}thi}, {Hudson}, \& {Gary}}]{Raulin2004}
{Raulin}, J.~P., {Makhmutov}, V.~S., {Kaufmann}, P., {et~al.} 2004, \solphys,
  223, 181, \dodoi{10.1007/s11207-004-1300-y}

\bibitem[{{Rubio da Costa} {et~al.}(2016){Rubio da Costa}, {Kleint},
  {Petrosian}, {Liu}, \& {Allred}}]{Rubio2016}
{Rubio da Costa}, F., {Kleint}, L., {Petrosian}, V., {Liu}, W., \& {Allred},
  J.~C. 2016, \apj, 827, 38, \dodoi{10.3847/0004-637X/827/1/38}

\bibitem[{{Sadykov} {et~al.}(2019){Sadykov}, {Kosovichev}, {Sharykin}, \&
  {Kerr}}]{Sadykov2019}
{Sadykov}, V.~M., {Kosovichev}, A.~G., {Sharykin}, I.~N., \& {Kerr}, G.~S.
  2019, \apj, 871, 2, \dodoi{10.3847/1538-4357/aaf6b0}

\bibitem[{{Shibayama} {et~al.}(2013){Shibayama}, {Maehara}, {Notsu}, {Notsu},
  {Nagao}, {Honda}, {Ishii}, {Nogami}, \& {Shibata}}]{Shibayama2013}
{Shibayama}, T., {Maehara}, H., {Notsu}, S., {et~al.} 2013, \apjs, 209, 5,
  \dodoi{10.1088/0067-0049/209/1/5}

\bibitem[{{Smith}(1975)}]{Smith1975}
{Smith}, D.~F. 1975, \apj, 201, 521, \dodoi{10.1086/153914}

\bibitem[{{Thorne} \& {Blandford}(2017)}]{Thorne2017}
{Thorne}, K.~S., \& {Blandford}, R.~D. 2017, {Modern Classical Physics: Optics,
  Fluids, Plasmas, Elasticity, Relativity, and Statistical Physics}

\bibitem[{{Tilley} {et~al.}(2019){Tilley}, {Segura}, {Meadows}, {Hawley}, \&
  {Davenport}}]{Tilley2019}
{Tilley}, M.~A., {Segura}, A., {Meadows}, V., {Hawley}, S., \& {Davenport}, J.
  2019, Astrobiology, 19, 64, \dodoi{10.1089/ast.2017.1794}

\bibitem[{{Tsiklauri}(2017)}]{Tsiklauri2017}
{Tsiklauri}, D. 2017, Physics of Plasmas, 24, 072902, \dodoi{10.1063/1.4990560}

\bibitem[{Tsytovich(1995)}]{Tsytovich1995}
Tsytovich, V.~N. 1995, Non-linear Plasma Kinetics in Modern Physics (Berlin,
  Heidelberg: Springer Berlin Heidelberg), 1--22,
  \dodoi{10.1007/978-3-642-78902-1_1}

\bibitem[{{van den Oord}(1990)}]{Oord1990}
{van den Oord}, G.~H.~J. 1990, \aap, 234, 496

\bibitem[{{Vilmer} {et~al.}(2011){Vilmer}, {MacKinnon}, \&
  {Hurford}}]{Vilmer2011}
{Vilmer}, N., {MacKinnon}, A.~L., \& {Hurford}, G.~J. 2011, \ssr, 159, 167,
  \dodoi{10.1007/s11214-010-9728-x}

\bibitem[{{Warmuth} {et~al.}(2009){Warmuth}, {Holman}, {Dennis}, {Mann},
  {Aurass}, \& {Milligan}}]{Warmuth2009}
{Warmuth}, A., {Holman}, G.~D., {Dennis}, B.~R., {et~al.} 2009, \apj, 699, 917,
  \dodoi{10.1088/0004-637X/699/1/917}

\bibitem[{{White} {et~al.}(2003){White}, {Krucker}, {Shibasaki}, {Yokoyama},
  {Shimojo}, \& {Kundu}}]{White2003}
{White}, S.~M., {Krucker}, S., {Shibasaki}, K., {et~al.} 2003, \apjl, 595,
  L111, \dodoi{10.1086/379274}

\bibitem[{{White} {et~al.}(2011){White}, {Benz}, {Christe}, {F{\'a}rn{\'\i}k},
  {Kundu}, {Mann}, {Ning}, {Raulin}, {Silva-V{\'a}lio}, {Saint-Hilaire},
  {Vilmer}, \& {Warmuth}}]{White2011}
{White}, S.~M., {Benz}, A.~O., {Christe}, S., {et~al.} 2011, \ssr, 159, 225,
  \dodoi{10.1007/s11214-010-9708-1}

\bibitem[{{Zharkova} \& {Gordovskyy}(2006)}]{Zharkova2006}
{Zharkova}, V.~V., \& {Gordovskyy}, M. 2006, \apj, 651, 553,
  \dodoi{10.1086/506423}

\end{thebibliography}
\bibliographystyle{aasjournal}

\end{document}